\documentclass[12pt]{article}
\usepackage{epsfig,amsmath,amssymb,mathrsfs}
\usepackage[utf8]{inputenc}
\usepackage[colorlinks=true,citecolor=blue,,linktocpage=true,linkcolor=blue,urlcolor=black]{hyperref}
\usepackage{mathtools,slashed}
\usepackage{array}
\usepackage{xcolor}
\usepackage[force]{feynmp-auto}
\usepackage{caption}
\usepackage{subcaption}
\usepackage{overpic,youngtab}
\usepackage[latin,english]{babel}
\usepackage{amsmath}
\usepackage{amssymb}

\usepackage{epstopdf}
\usepackage{graphics,psfrag,rotating}
\usepackage{mathabx}
\usepackage{graphicx}
\usepackage{dcolumn}
\usepackage{float}
\usepackage{pdflscape}
\usepackage{array}
\usepackage{booktabs}
\usepackage{amscd}
\usepackage{mathtools}
\usepackage{fancybox}
\usepackage{fix-cm}
\usepackage[colorlinks=true,citecolor=blue,,linktocpage=true,linkcolor=blue,urlcolor=black]{hyperref}
\usepackage{tikz}
\usetikzlibrary{matrix}
\usetikzlibrary{positioning}
\usetikzlibrary{arrows}

  {\end{list}}%

\makeatletter
\renewcommand{\section}{\setcounter{equation}{0}\@startsection
 {section}%
 {1}%
 {0pt}%
 {-1\baselineskip}%
 {0.4\baselineskip}%
 {\bfseries\large}}%
\renewcommand{\subsection}{\@startsection
 {subsection}%
 {2}%
 {0pt}%
 {-0.75\baselineskip}%
 {0.2\baselineskip}%
 {\bfseries}}%
\renewcommand{\subsubsection}{\@startsection
 {subsubsection}%
 {3}%
 {0pt}%
 {-0.5\baselineskip}%
 {0.1\baselineskip}%
 {\bfseries}}%
\makeatother

\DeclareMathAlphabet{\mathpzc}{OT1}{pzc}{m}{it}

\usepackage{tikz}
\usetikzlibrary{patterns,snakes}
\usetikzlibrary{shapes.misc}
\usetikzlibrary{decorations.markings,decorations.pathmorphing}
\usetikzlibrary{calc}
\tikzstyle{spring}=[line width=0.8,black,snake=coil,segment amplitude=4.25,segment length=4.75,line cap=round]

\def\be{\begin{equation}}
\def\ee{\end{equation}}

\def\g5{\gamma_{5}}

\def\Dirac{{D\mkern-12mu/}}

\def\pslash{{p\mkern-8mu/}{\!}}

\def\id3k{\int\!\! \dfrac{d^3\!\vec{k}}{(2\pi)^3 }}
\def\idp{\int\!\! \dfrac{d^4\!p}{(2\pi)^4}}

\def\dpi{\dfrac{d^4\!p_j}{(2\pi)^4}}

\def\idq{\int\!\! \dfrac{d^4\!q}{(2\pi)^4} \,\,}

\def\idqD{\int\!\! \dfrac{d^{D}q\!}{(2\pi)^{D}}}

\def\idx{\int\!\! d^4\!x}

\def\idx{\int d^{4}\!x}

\newcommand{\bea}{\begin{eqnarray}}
\newcommand{\eea}{\end{eqnarray}}
\newcommand{\beann}{\begin{eqnarray*}}
\newcommand{\eeann}{\end{eqnarray*}}
\newcommand{\ba}{\begin{array}}
\newcommand{\ea}{\end{array}}

\newcommand{\tr}{\mathbf{tr}}

 \def\g {\gamma}

\newcommand{\email}[1]{\href{mailto:#1}{\tt #1}}

\begin{document}

\rightline{\scriptsize{}}
\vglue 50pt

\begin{center}

{\LARGE \bf The effective gravitational action of a massless chiral fermion and the absence of parity-odd contributions.}\\
\vskip 1.0true cm
{\Large Jes\'us Anero$^{\dagger} $and Carmelo P. Mart\'{\i}n$^{\dagger\dagger}$ }
\\
\vskip .7cm
{
	$\dagger$Departamento de F\'isica Te\'orica and Instituto de F\'{\i}sica Te\'orica (IFT-UAM/CSIC),\\
	Universidad Aut\'onoma de Madrid, Cantoblanco, 28049, Madrid, Spain\\
	\vskip .1cm
	{$\dagger\dagger$Universidad Complutense de Madrid (UCM), Departamento de Física Teórica and IPARCOS, Facultad de Ciencias Físicas, 28040 Madrid, Spain}
	
	\vskip .5cm
	\begin{minipage}[l]{.9\textwidth}
		\begin{center}
			\textit{E-mail: jesusanero@gmail.com, carmelop@fis.ucm.es}
			\email{}
			
		\end{center}
	\end{minipage}
}
\end{center}
\thispagestyle{empty}

\begin{abstract}
We consider the field theory of a quantum massless left handed fermion  coupled to a background graviton field, $h_{\mu\nu}$, on  four-dimensional Minkowski spacetime.
By using the BPHZL renormalization scheme, we prove that, up to order four in the number of graviton fields, there are no parity-odd contributions to  the renormalized gravitational effective action. As a side result, we show that, modulo arbitrary UV finite  diffeomorphism invariant parity-even counterterms, the gravitational effective action in question is equal, up to order four in the number of graviton fields, to half the corresponding  gravitational action for a Dirac fermion non-chirally coupled to gravity. Also as a side result, we conclude the Weyl anomaly is purely parity-even and that its value is half the value of the Weyl anomaly for a Dirac fermion non-chirally coupled to gravity.

\end{abstract}

\vfill
\clearpage

\section{Introduction}

In four dimensional spacetime, the gauge groups $U(1)$, $SU(N)$ and the group of diffeomorphisms play a fundamental role in the description of Nature.
 It is well known \cite{Alvarez-Gaume:1985zzv} that, generally speaking, the way  Weyl fermions interact at quantum level with $U(1)$ and $SU(N)$ gauge fields, on the one hand, and the graviton field, on the other hand, is quite different. Indeed, gauge anomalies may arise in the first case, whereas no diffeomorphism anomaly develops in four dimensions when  Weyl fermions are coupled to the gravitational field.

In a celebrated paper \cite{Coleman:1982yg}, Coleman and Grossman showed that if the anomaly cancellation conditions are not met by the fermion representations the one-loop fermion contribution to the gauge field three-point function develops a pole whose residue is a parity-odd polynomial in the momenta. This pole structure, which is parity-odd and UV finite,  gives rise to the gauge anomaly and goes away when the anomaly cancellation conditions are satisfied. Hence, the fact that there are no purely gravitational anomalies in four dimensions raises  the question  as to whether the renormalized three-point part of the  gravitational effective action carries any one-loop parity-odd contributions coming from  Weyl fermions running along the loop. In this regard let us point out that the purely cohomological analysis in \cite{Bonora:1985cq} does not preclude the existence of a parity-odd contribution to the Weyl anomaly and, hence, the existence of parity-odd contributions to the gravitational effective action. So it seems that only by explicit computation the question of whether there are parity-odd contributions -local and non-local-- to the gravitational effective action can be settled.\footnote{The explicit computations in \cite{Bastianelli:2016nuf, Abdallah:2021eii, Larue:2023tmu, Alvarez:2025mql} yield the result that there is no parity-odd contribution to the Weyl anomaly in four  dimensions, which utterly contradicts the earlier non-vanishing result in \cite{Bonora:2014qla, Bonora:2023soh}.}

There is a sizeable amount of interest in the study of the violation of parity in the gravitational field by scrutinizing their effects on gravitational waves --see \cite{Zhu:2023rrx}, \cite{Chakraborty:2025qcu} and references therein. In this regard it is of significant relevance to ascertain whether that parity breaking is generated by quantum interaction with chiral matter. Of course, arguably, Weyl fermions fall into the category of the most important type of chiral matter.

In view of the previous considerations, we think that it is interesting to consider a single massless left-handed fermion coupled to the graviton field and  find out whether there are parity breaking contributions to the effective action of the graviton field on the Minkowski background. We shall show below that there are no such parity-odd  contributions to the three and four-point functions of the renormalized gravitational action. Further, our computations yield two side results, which are valid at least up to order four in the number of graviton fields. First, that the parity-even part of effective action is, modulo arbitrary diffeomorphism invariant counterterms,  half the gravitational effective action for a quantum Dirac field non-chirally coupled to gravity. Secondly, that we agree with references  \cite{Bastianelli:2016nuf, Abdallah:2021eii, Larue:2023tmu} and \cite{Alvarez:2025mql} in that there is no parity-odd  contribution to the Weyl anomaly for the theory under scrutiny.

To show that, at one-loop, the one, two, three and four-point  contributions to the gravitational effective action  of a background graviton field coupled to a single massless left  handed spinor carry no parity-odd contributions, we shall define the gravitational effective action by using a massless Dirac spinor chirally coupled to the graviton field. We shall regularize and renormalize  such effective action by employing  the Pauli-Villars regularization and the BPHZL renormalization procedures formulated in \cite{Clark:1976ym} --see also \cite{Lowenstein:1974qt, Lowenstein:1975ps, Lowenstein:1975ug}. These procedures constitute a refinement of the original regularization plus renormalization method  in \cite{Bogolyubov:1959bfo, Zimmermann:1975gk}, for the latter cannot be applied to massless theories.

As is well known the regularization plus renormalization procedure in \cite{Clark:1976ym}  is a rigorous way of constructing renormalized perturbative field theory in Minkowski  spacetime, whose chief virtue is that it satisfies the action principle. Hence, any one-loop breaking of a classical gauge symmetry the procedure gives rise to shows as a local  operator; and, thus, the breaking in question can always be removed by adding a suitable local  compensating counterterm, if  the symmetry is not anomalous.

The Pauli-Villars regularization plus the BPHZL renormalization procedure in \cite{Clark:1976ym} clearly breaks the classical diffeomorphism invariance of our theory. However, it is well known that the diffeomorphism invariance of this theory is not anomalous (we are in 4 dimensions) and, hence, as discussed in the previous paragraph, the invariance under diffeomorphisms can always be restored by introducing appropriate counterterms.  What is key is that these counterterms up to order four (included) in the number of fields have no parity-odd terms, for we shall show that the regularized gravitational effective action defined as in \cite{Clark:1976ym} carries no parity-odd contributions up to four in the number of fields.

The layout of this paper is as follows. In section 2, we put forward the model, giving its classical action and defining, formally, its partition function by introducing
a non-interacting right-handed fermion. The Pauli-Villars regularization --in the sense of \cite{Clark:1976ym}-- of the model is presented in section 3. In section 4, we show that the regularized gravitational effective action of the model has no parity-odd contributions up to order four in the number of graviton fields. The way to apply the BPHZL renormalization algorithm to our regularized effective action is discussed in section 5. In this section 5, we show that the renormalized effective gravitational action
does not contain parity-odd contributions, up to order four in the number of graviton fields, and that its value is half the value of the corresponding dimensionally  renormalized gravitational effective action of a Dirac field non-chirally coupled to the graviton field on Minkowski spacetime. The use of higher derivative regularizations to draw the same conclusions as with the  Pauli-Villars regularization of  Ref.~\cite{Clark:1976ym} is analyzed in section 6. In section 7, we tackle a final discussion of the results obtained in the previous  sections. In appendix A we give  a brief summary of how the computations in the paper have been done. Appendix B is devoted to the discussion of how to restore the broken diffeomorphism invariance by introducing UV finite counterterms.

\newpage
\section{The model, its classical action and the effective action.}

Let $\omega_{\mu a b}(x)$ be the spin connection
\begin{equation*}
\omega_{\mu a b}(x)= -e^{\lambda}_{b}(x)\partial_\mu e_{\lambda a}(x)+e_{\lambda a}(x)\Gamma^{\lambda}_{\rho \mu}(x)e^{\rho}_{b}(x).
\end{equation*}
Then, the classical action of the model reads
\begin{equation*}
S=\cfrac{1}{2}\,\idx\,e(x)\,e^{\mu}_{a}(x)\,[i\bar{\psi}_L(x)\gamma^{ a}({\cal D}_{\mu}\psi_{L})(x)-i
({\cal D}_\mu\bar{\psi}_L)(x)\gamma^{ a}\psi_{L}(x)],
\end{equation*}
where
\begin{equation*}
{\cal D}_{\mu}\psi_{L}(x)=\partial_\mu\psi_{L}(x)-\cfrac{i}{2}\omega_{\mu a b}(x)\,\gamma^{a b}\psi_{L}(x),\quad
{\cal D}_\mu\bar{\psi}_{L}(x)=\partial_{\mu}\bar{\psi}_{L}(x)+\cfrac{i}{2}\bar{\psi}_{L}(x)\gamma^{ab}\,\omega_{\mu a b}(x).
\end{equation*}

Since we shall construct the quantum theory for the  fermions in perturbation theory around Minkowski spacetime, we shall express the metric, $g_{\mu\nu}$, the vierbein $e^a_\mu$, the inverse vierbein, $e^{\mu}_a$, and the spin connection, $\omega_{a, bc}$ as follows
\begin{equation*}
\begin{array}{l}
{g_{\mu\nu}=\eta_{\mu\nu}+\kappa h_{\mu\nu},}\\[4pt]
{e^a_{\mu}=\delta^a_\mu+\kappa{\cal E}^{(1)\,a}_{\mu}+\kappa^2 {\cal E}^{(2)\,a}_{\mu}+...}\\[4pt]
{e^\mu_{a}=\delta^\mu_a+\kappa{\cal E}^{(1)\,\mu}_{a}+\kappa^2 {\cal E}^{(2)\,\mu}_{a}+...}\\[4pt]
{\omega_{a, bc}=e^{\mu}_a\omega_{\mu bc}=\kappa \Omega^{(1)}_{a, bc}+\kappa^2 \Omega^{(2)}_{a, bc}+....,}
\end{array}
\end{equation*}
where
\begin{equation*}
\begin{array}{l}
{{\cal E}^{(1)\,a}_{\mu}=\cfrac{1}{2}h_a^\mu,\quad {\cal E}^{(2)\,a}_{\mu}=-\frac{1}{8}\,h^{a\rho}h_{\rho}^{\mu},\quad
{\cal E}^{(1)\,\mu}_{a}=-\cfrac{1}{2}h^\mu_a,\quad {\cal E}^{(2)\,\mu}_{a}=\frac{3}{8}\,h^{\mu\rho}h_{\rho a}}\\[4pt]
{\Omega^{(1)}_{a, bc}=-\frac{1}{2}\,(\partial_b h_{ca}-\partial_c h_{b a})}\\[4pt]
{\Omega^{(2)}_{a, bc}=\frac{1}{4}h^{\mu}_a(\partial_b h_{c\mu}-\partial_c h_{b\mu})-\frac{1}{8}(h^{\rho}_b\partial_a h_{c\rho}-
h^{\rho}_c\partial_a h_{b\rho})+\frac{1}{4}(h^{\rho}_b\partial_\rho h_{c a}-h^{\rho}_c\partial_\rho h_{b a})-\frac{1}{4}
(h^{\rho}_b\partial_c h_{\rho a}-h^{\rho}_c\partial_b h_{\rho a}).}
\end{array}
\end{equation*}

The previous expansions in powers of $\kappa$, lead to the following expansion of the action $S$:
\begin{equation}
\begin{array}{l}
{S=S_{0}+S_{int}}\\[4pt]
{S_{0}=\,\idx\,\,i\bar{\psi}_{L}(x)\gamma^{\mu}\partial_{\mu}\psi_L(x),}\\[4pt]
{S_{int}=\sum\limits_n\,\kappa^n S^{(n)}.}
\end{array}
\label{Laction}
\end{equation}
In both $S_0$ and $S_{int}$ all the indices are flat. $S_0$ is the free action in Minkowski spacetime and $S_{int}$ carries the interaction vertices between the graviton field $h_{\mu\nu}$ and the fermion $\psi_L$. $S^{(n)}$ is quadratic on the fermion fields and involves $n$ powers of $h_{\mu\nu}$ and/or its derivatives. A little computation yields
\begin{equation}
\begin{array}{l}
{S_{int}=\,S_{kin}+S_{spin},}\\[4pt]
{S_{kin}=\idx\,
\frac{i}{2}\,H^{\mu}_{\nu}(x)\,[\bar{\psi}_L(x)\gamma^{\nu}\partial_{\mu}\psi_L(x)-
\partial_\mu\bar{\psi}_L(x)\gamma^{\nu}\psi_L(x)],}\\[4pt]
{S_{spin}=\idx\,\epsilon^{\mu\nu\rho\sigma}\,H_{\mu\nu\rho}(x)\,
\bar{\psi}_L(x)\gamma_\sigma\psi_L(x),}
\end{array}
\label{Sintexpan}
\end{equation}
where
\begin{equation}
\begin{array}{l}
{H^{\mu}_{\nu}(x)=\kappa H^{\mu}_{1,\,\nu}(x)+\kappa^2\, H^{\mu}_{2,\,\nu}(x)+\cdots}\\[4pt]
{H_{\mu\nu\rho}(x)=\kappa^2\,H_{2,\mu\nu\rho}(x)+\cdots}\\[4pt]
{H^{\mu}_{1,\,\nu}(x)=\frac{1}{2}\,(h^{\rho}_{\rho}(x)\delta^{\mu}_{\nu}-h^{\mu}_{\nu}(x)),}\\[4pt]
{H^{\mu}_{2,\,\nu}(x)=-\frac{1}{4}h^{\rho}_{\rho}(x)h^{\mu}_{\nu}(x)+\frac{1}{8}(h^{\rho}_{\rho}(x))^2\delta^{\mu}_{\nu}-\frac{1}{4}
h^{\rho\sigma}(x)h_{\rho\sigma}(x)+\frac{3}{8}h^{\mu\rho}(x)h_{\rho \nu}(x),}\\[4pt]
{H_{2,\mu\nu\rho}(x)=-\frac{1}{16}\,h^\lambda_{\mu}(x)\partial_{\nu}h_{\rho\lambda}(x).}
\end{array}
\label{Hexpansion}
\end{equation}
As we shall see the  actual value of $H^{\mu}_{\nu}(x)$ and $H_{\mu\nu\rho}(x)$  will not be needed. The only features of  $H^{\mu}_{\nu}(x)$ and $H_{\mu\nu\rho}(x)$ we shall use is that  they only depend on $h_{\mu\nu}$ and its derivatives, and that $H^{\mu}_{\nu}(x)$ is order one in $\kappa$ whereas
$H_{\mu\nu\rho}(x)$ is order 2 in $\kappa$. Of course, all the indices in (\ref{Hexpansion}) are flat.

Let  ${\cal Z}[h_{\mu\nu}]$ denote the partition function of the model. To define it in perturbation theory around a Minkowski background, we shall introduce a spectator right-handed fermion field --i.e., a  right-handed fermion which does not couple to $h_{\mu\nu}$, and replace the action $S$ in  (\ref{Laction}) with ${\cal S}$ defined as follows:
\begin{equation}
\begin{array}{l}
{{\cal S}={\cal S}_{0}+S_{int}}\\[4pt]
{{\cal S}_{0}=\,\idx\,\,i\bar{\psi}(x)\gamma^{\mu}\partial_{\mu}\psi(x),}
\end{array}
\label{action}
\end{equation}
where $S_{int}$ is defined in (\ref{Sintexpan}) and (\ref{Hexpansion}).

In the previous definition $\psi$ is the Dirac fermion defined by
\begin{equation*}
\begin{array}{l}
{\psi(x)=\psi_{L}(x)+\psi_{R}(x),\psi_L(x)=P_{L}\psi(x),\psi_R(x)=P_{R}\psi(x),}\\[4pt]
{P_L=\cfrac{1}{2}(1-\gamma_5),  P_R=\cfrac{1}{2}(1+\gamma_5), \gamma_5=i\gamma^0\gamma^1\gamma^2
\gamma^3.}
\end{array}
\end{equation*}
Let us stress that $S_{int}$ in {\ref{action}) is the one in (\ref{Sintexpan}), so that  $\psi_R$ does not occur  in it.

Then, ${\cal Z}[h_{\mu\nu}]$ is formally defined as follows
 \begin{equation*}
\begin{array}{l}
{{\cal Z}[h_{\mu\nu}]=\cfrac{1}{{\cal N}}\int \mathscr{D}\psi\mathscr{D}\bar{\psi}\;e^{i{\cal S}}=\langle\,e^{iS_{int}}\,\rangle_0}\\[4pt]
{{\cal N}=\int \mathscr{D}\psi\mathscr{D}\bar{\psi}\;e^{i{\cal S}_{0}}.}
\end{array}
\end{equation*}
$\langle {\cal O}\rangle_{0}$ denotes the average of ${\cal O}$  with regard to the free action ${\cal S}_{0}$ in (\ref{action}):
\begin{equation*}
\langle {\cal O}\rangle_{0}=\cfrac{1}{{\cal N}}\int \mathscr{D}\psi\mathscr{D}\bar{\psi}\;{\cal O}\;e^{i{\cal S}_0}
\end{equation*}

Then, the formal effective action, ${\cal W}[h_{\mu\nu}]$, of the model is defined to be
\begin{equation}
{\cal W}[h_{\mu\nu}]= -i\, {\rm Ln} {\cal Z}[h_{\mu\nu}]=-i\sum_n\,\cfrac{i^n}{n!}\,\langle (S_{int})^n\rangle_{0}^{(c)},
\label{Qaction}
\end{equation}
The superscript "c" signals that only connected contractions, as given by Wick's theorem, are to be kept.  Note that in the case at hand the free fermion propagator entering the Wick's theorem reads
\begin{equation}
\langle \psi_{\alpha}(x)\bar{\psi}_{\beta}(y)\rangle_0=\idp\;\pslash_{\alpha\beta}\;K(p)\;e^{-ip(x-y)},\quad
K(p)=\cfrac{i}{p^2}.
\label{freepropnonreg}
\end{equation}

\section{The  regularized effective action.}

As it stands, ${\cal W}[h_{\mu\nu}]$, in (\ref{Qaction}) is an ill-defined object due to UV divergences. As announced in the introduction, to regularize these UV divergences we shall use the
Pauli-Villars method as formulated in  \cite{Clark:1976ym} in connection with the action principle. Hence, when carrying out the Wick's contractions we shall
replace the unregularized propagator in (\ref{freepropnonreg}) with the regularized one
\begin{equation}
\begin{array}{l}
{\langle \psi_{\alpha}(x)\bar{\psi}_{\beta}(y)\rangle_0=\idp\; [\pslash_{\alpha\beta}+ (s-1) \,m\,\delta_{\alpha\beta}]\,G^{(reg)}(p)\;e^{-ip(x-y)},}\\[8pt]
{G^{(reg)}(p)=\cfrac{i}{p^2-(s-1)^2 m^2+i\varepsilon(\vec{p}^2+m^2(s-1)^2)}+\sum\limits_{i=1}^n\,c_i\,\cfrac{i}{p^2-M_i^2+i\varepsilon(\vec{p}^2+M_i^2)},}
\end{array}
\label{freeprop}
\end{equation}
where  $M_i$ and $c_i$ are the masses and Pauli-Villars parameters  and  $(s-1) m$, $s\in[0,1]$, is the auxiliary mass needed to carry out the appropriate IR subtractions. The reader is referred to reference  \cite{Clark:1976ym}  for further details. Let us point
out that regarding our computations, presented in the sequel, one only needs to know that such regularization exists and that the BPHZL renormalization process based on it yields a renormalized effective action which satisfies the action principle. Of course, the $c_i$'s
and $M_i$'s above must be chosen so that the Feynman diagrams are regularized, {\it i.e.}, so that the  $G^{(reg)}(p)$ goes to zero as $p^{-2n}$, $n\geq3$, for large $p$, as some trivial power-counting tells.

 We should like to point out that the  Pauli-Villars regularization scheme introduced above is not the Pauli-Villars method employed in \cite{Alvarez:2025mql} to conclude that there is no parity-odd contribution to the Weyl  anomaly due to Weyl Fermions.
Indeed, the Pauli-Villars regularization in (\ref{freeprop}) improves the UV behaviour of each free propagator individually, this leads to a breaking of diffeomorphism invariance. On the other hand, The Pauli-Villars scheme used in \cite{Alvarez:2025mql} deals with the fermion loops as a whole: one regularizes each loop by considering an appropriate linear combination of  the loop contribution coming from the physical field and the  loop contributions due to Pauli-Villars fields. This way the Pauli-Villars method in \cite{Alvarez:2025mql} preserves diffeomorphism invariant explicitly. Notice that the regularization method used in  \cite{Alvarez:2025mql} is an adaptation of the Pauli-Villars regularization method used to regularize the fermion loop contributions in QED --see, for instance, \cite{Itzykson:1980rh}.

The reader should bear in mind that although the masses in the propagator in (\ref{freeprop}) are Dirac, $S_{int}$ in (\ref{action}) is not modified in our regularization method, so that the interaction between the graviton field and the fermion fields is chiral. Then, the regularized effective action,
${\cal W}^{(PV)}[h_{\mu\nu}]$, is defined as follows
 \begin{equation}
 {\cal W}^{(PV)}[h_{\mu\nu}]=-i\sum_n\,\cfrac{i^n}{n!}\,\langle (S_{int})^n\rangle_{0}^{(PV,\,c\,)}=\sum_{n=1}^\infty\,\kappa^n\,
 {\cal W}^{(PV)}_n[h_{\mu\nu}],
 \label{PVQaction}
 \end{equation}
 where  $\langle (S_{int})^n\rangle_{0}^{(PV,\,c\,)}$ are the connected contributions obtained by applying the Wick's theorem for the regularized  free propagators in (\ref{freeprop}).

\section{No parity-odd contributions to  ${\cal W}^{(PV)}[h_{\mu\nu}]$ up to order 4 in the number of $h_{\mu\nu}$.}

The contribution to the effective action, ${\cal W}^{(PV)}[h_{\mu\nu}]$, involving up to four $h_{\mu\nu}$ reads
\begin{equation*}
{\cal F}^{(PV)}[h_{\mu\nu}]=\sum_{n=1}^4\,\kappa^n\,
 {\cal W}^{(PV)}_n[h_{\mu\nu}].
\end{equation*}

Taking into account (\ref{Sintexpan}), (\ref{Hexpansion}) and (\ref{PVQaction}), one concludes ${\cal F}^{(PV)}[h_{\mu\nu}]$ above can be obtained by
extracting from
\begin{equation}
\begin{array}{l}
{{\cal G}^{(PV)}[h_{\mu\nu}]=\langle S_{kin}\rangle_{0}^{(PV,\,c\,)}+\frac{1}{2}\,i\,\langle S_{kin}^2\rangle_{0}^{(PV,\,c\,)}\,+\, \langle S_{spin}\rangle_{0}^{(PV,\,c\,)}-\frac{1}{6}\, \langle S_{kin}^3\rangle_{0}^{(PV,\,c\,)}+}\\[4pt]
{\quad\quad i\, \langle S_{kin}S_{spin}\rangle_{0}^{(PV,\,c\,)}- \frac{1}{2} \langle S_{kin}^2 S_{spin}\rangle_{0}^{(PV,\,c\,)}
+\frac{i}{2}\,\langle S_{spin}^2\rangle_{0}^{(PV,\,c\,)}-\frac{i}{24}\,\langle  S_{kin}^4\rangle_{0}^{(PV,\,c\,)}}
\end{array}
\label{GPV}
\end{equation}
the contributions involving one, two, three and four fields $h_{\mu\nu}$.

Hence, by showing that there are no parity-odd contributions to
${\cal G}^{(PV)}[h_{\mu\nu}]$, we will show that there are no parity-odd contributions to ${\cal F}^{(PV)}[h_{\mu\nu}]$, ie, no parity-odd contributions to the effective action ${\cal W}^{(PV)}[h_{\mu\nu}]$ up to order four in the number of fields.

Before we display our results let us stress that in obtaining them  the following properties of the Dirac matrices in four dimensions have been extensively used
\begin{equation}
\begin{array}{l}
{\tr[\gamma^{\mu_1}\cdots\gamma^{\mu_{2n+1}}]=0,}\\[8pt]
{\tr[\gamma^{\mu_{1}}\cdots\gamma^{\mu_{2n}}]=\tr[\gamma^{\mu_{2n}}\cdots\gamma^{\mu_{1}}],}\\[8pt]
{P_L\gamma^{\mu}P_L=0.}
\end{array}
\label{DiracM}
\end{equation}

\subsection{No parity-odd contributions to $\langle S_{kin}\rangle_{0}^{(PV,\,c\,)}$.}
The contribution
\begin{equation*}
\langle S_{kin}\rangle_0^{(PV,\,c\,)}=\frac{i}{2}\,\int\,d^4x\,H_{\nu}^\mu(x)\,\langle[\bar{\psi}(x)\gamma^{\nu}P_L\partial_\mu\psi(x)-
\partial_\mu\bar{\psi}(x)\gamma^{\nu}P_L\psi(x)]\rangle_0
\end{equation*}
carries no parity-odd term since one needs, in the least, to trace over the product of four Dirac matrices and one $\gamma_5$ to generate a Levi-Civita pseudotensor.

\subsection{No parity-odd contribution to $\langle S_{kin}^2\rangle_{0}^{(PV,\,c\,)}$.}

It can be seen that
\begin{equation}
\langle S_{kin}^2\rangle_0^{(PV,\,c\,)} =-\cfrac{1}{4}\,\big({\cal A}_1+{\cal A}_2+{\cal A}_3+{\cal A}_4\big),
\label{selfenkin}
\end{equation}
where
\begin{equation*}
\begin{array}{l}
{{\cal A}_1=\int dx_1\,dx_2\,H^{\mu_1}_{\nu_1}(x_1)\,H^{\mu_2}_{\nu_2}(x_2)\,
\langle\bar{\psi}_L(x_1)\gamma^{\nu_1}\partial_{\mu_1}\psi_L(x_1)\bar{\psi}_L(x_2)\gamma^{\nu_2}\partial_{\mu_2}\psi_L(x_2)\rangle_0^{(PV,\,c\,)},}\\[12pt]

{{\cal A}_2=\int dx_1\,dx_2\,H^{\mu_1}_{\nu_1}(x_1)\,H^{\mu_2}_{\nu_2}(x_2)\,
\langle\partial_{\mu_1}\bar{\psi}_L(x_1)\gamma^{\nu_1}\psi_L(x_1)\partial_{\mu_2}\bar{\psi}_L(x_2)\gamma^{\nu_2}\psi_L(x_2)\rangle_0^{(PV,\,c\,)},}\\[12pt]

{{\cal A}_3=-\int dx_1\,dx_2\,H^{\mu_1}_{\nu_1}(x_1)\,H^{\mu_2}_{\nu_2}(x_2)\,
\langle\bar{\psi}_L(x_1)\gamma^{\nu_1}\partial_{\mu_1}\psi_L(x_1)\partial_{\mu_2}\bar{\psi}_L(x_2)\gamma^{\nu_2}\psi_L(x_2)\rangle_0^{(PV,\,c\,)},}\\[12pt]

{{\cal A}_4=-\int dx_1\,dx_2\,H^{\mu_1}_{\nu_1}(x_1)\,H^{\mu_2}_{\nu_2}(x_2)\,
\langle\partial_{\mu_1}\bar{\psi}_L(x_1)\gamma^{\nu_1}\psi_L(x_1)\bar{\psi}_L(x_2)\gamma^{\nu_2}\partial_{\mu_2}\psi_L(x_2)\rangle_0^{(PV,\,c\,)},}\\[12pt]

\end{array}
\end{equation*}

By using Wick's theorem and the properties of Dirac matrices, one shows that

\begin{equation}
\begin{array}{l}
{{\cal A}_1={\cal A}_1^{(even)}+{\cal A}_1^{(odd)}}\\[8pt]
{{\cal A}_1^{(even)}= \cfrac{1}{2}\int\prod\limits_{i=1}^{2}\dpi(2\pi)^4\,\delta(p_1+p_2)\,H^{\mu_1\nu_1}(p_1)\,H^{\mu_2\nu_2}(p_2)\Big\{
\tr[\gamma_{\nu_1}\gamma_{\sigma_1}\gamma_{\nu_2}\gamma_{\sigma_2}]}\\[4pt]
{\quad\quad\quad\idq
\,q_{\mu_1}q^{\sigma_1}(q+p_2)_{\mu_2}(q+p_2)^{\sigma_2}\,G^{(reg)}(q)\,G^{(reg)}(q+p_2)\Big\},}\\[12pt]

{{\cal A}_1^{(odd)}= \cfrac{1}{2}\int\prod\limits_{i=1}^{2}\dpi(2\pi)^4\,\delta(p_1+p_2)\,H^{\mu_1\nu_1}(p_1)\,H^{\mu_2\nu_2}(p_2)\Big\{
\tr[\gamma_{\nu_1}\gamma_{\sigma_1}\gamma_{\nu_2}\gamma_{\sigma_2}\gamma_5]}\\[4pt]
{\quad\quad\quad  \idq
\, q_{\mu_1}q^{\sigma_1}(q+p_2)_{\mu_2}(q+p_2)^{\sigma_2}
\,G^{(reg)}(q)\,G^{(reg)}(q+p_2)
\Big\},}\\[20pt]

{{\cal A}_2={\cal A}_2^{(even)}+{\cal A}_2^{(odd)}}\\[8pt]
{{\cal A}_2^{(even)}={\cal A}_1^{(even)},\quad{\cal A}_2^{(odd)}=-{\cal A}_1^{(odd)},}\\[20pt]

{{\cal A}_3={\cal A}_3^{(even)}+{\cal A}_3^{(odd)}}\\[8pt]
{{\cal A}_3^{(even)}= \cfrac{1}{2}\int\prod\limits_{i=1}^{2}\dpi(2\pi)^4\,\delta(p_1+p_2)\,H^{\mu_1\nu_1}(p_1)\,H^{\mu_2\nu_2}(p_2)\Big\{
\tr[\gamma_{\nu_1}\gamma_{\sigma_1}\gamma_{\nu_2}\gamma_{\sigma_2}]}\\[4pt]
{\quad\quad\quad\idq
\,q_{\mu_1}q_{\mu_2}q^{\sigma_1}(q+p_2)^{\sigma_2}\,G^{(reg)}(q)\,G^{(reg)}(q+p_2)\Big\},}\\[12pt]

{{\cal A}_3^{(odd)}= \cfrac{1}{2}\int\prod\limits_{i=1}^{2}\dpi(2\pi)^4\,\delta(p_1+p_2)\,H^{\mu_1\nu_1}(p_1)\,H^{\mu_2\nu_2}(p_2)\Big\{
\tr[\gamma_{\nu_1}\gamma_{\sigma_1}\gamma_{\nu_2}\gamma_{\sigma_2}\gamma_5]}\\[4pt]
{\quad\quad\quad \idq
\, q_{\mu_1}q_{\mu_2}q^{\sigma_1}(q+p_2)^{\sigma_2}\,G^{(reg)}(q)\,G^{(reg)}(q+p_2)\Big\},}\\[20pt]

{{\cal A}_4={\cal A}_4^{(even)}+{\cal A}_4^{(odd)}}\\[8pt]
{{\cal A}_4^{(even)}={\cal A}_3^{(even)},\quad{\cal A}_4^{(odd)}=-{\cal A}_3^{(odd)},}\\[20pt]

\end{array}
\label{selfenerkinvalues}
\end{equation}

Combining (\ref{selfenkin}) and (\ref{selfenerkinvalues}), one gets
\begin{equation*}
\langle S_{kin}^2\rangle_0^{(PV,\,c\,)} =-\cfrac{1}{2}\,\big({\cal A}_1^{(even)}+{\cal A}_3^{(even)}\big).
\end{equation*}
No parity-odd contribution occurs here.

\subsection{No parity-odd contribution from  $\langle S_{spin}\rangle_0^{(PV,\,c\,)}$.}

The fact that the integration of a regularized odd integrand is zero leads to
\begin{equation*}
\langle S_{spin}\rangle_0^{(PV,\,c\,)}=\int\,d^4x\,H_{\mu\nu\rho}(x)\,\epsilon^{\mu\nu\rho\lambda}\,\langle\bar{\psi}(x)\gamma_{\lambda}P_L\psi(x)\rangle_0=
0.
\end{equation*}

\subsection{No parity-odd contribution from $\langle (S_{kin})^3\rangle_0^{(PV,\,c\,)}$.}

Taking into account the definitions in (\ref{Sintexpan}), one concludes that
\begin{equation}
\langle (S_{kin})^3\rangle_0^{(c)}=-\cfrac{i}{8}\,\big({\cal T}_1+{\cal T}_2+{\cal T}_3+{\cal T}_4\big),
\label{triangle}
\end{equation}
where
\begin{equation}
\begin{array}{l}
{{\cal T}_1=\int dx_1\,dx_2\,dx_3\,H^{\mu_1}_{\nu_1}(x_1)\,H^{\mu_2}_{\nu_2}(x_2)H^{\mu_3}_{\nu_3}(x_3)}\\[4pt]
{
\phantom{{\cal T}_1=\int dx_1\,dx_2\,dx_3\,,H}
\langle\bar{\psi}_L(x_1)\gamma^{\nu_1}\partial_{\mu_1}\psi_L(x_1)\bar{\psi}_L(x_2)\gamma^{\nu_2}\partial_{\mu_2}\psi_L(x_2)\bar{\psi}_L(x_3)\gamma^{\nu_3}
\partial_{\mu_3}\psi_L(x_3)\rangle_0^{(PV,\,c\,)},}\\[12pt]
{{\cal T}_2=-\int dx_1\,dx_2\,dx_3\,H^{\mu_1}_{\nu_1}(x_1)\,H^{\mu_2}_{\nu_2}(x_2)H^{\mu_3}_{\nu_3}(x_3)}\\[4pt]
{
\phantom{{\cal T}_2=-\int dx_1\,dx_2\,dx_3\,,H}
\langle\partial_{\mu_1}\bar{\psi}_L(x_1)\gamma^{\nu_1}\psi_L(x_1)\partial_{\mu_2}\bar{\psi}_L(x_2)\gamma^{\nu_2}\psi_L(x_2)\partial_{\mu_3}\bar{\psi}_L(x_3)\gamma^{\nu_3}
\psi_L(x_3)\rangle_0^{(PV,\,c\,)},}\\[12pt]

{{\cal T}_3=-3\int dx_1\,dx_2\,dx_3\,H^{\mu_1}_{\nu_1}(x_1)\,H^{\mu_2}_{\nu_2}(x_2)H^{\mu_3}_{\nu_3}(x_3)}\\[4pt]
{
\phantom{{\cal T}_3=-3\int dx_1\,dx_2\,dx_3\,,H}
\langle\bar{\psi}_L(x_1)\gamma^{\nu_1}\partial_{\mu_1}\psi_L(x_1)\bar{\psi}_L(x_2)\gamma^{\nu_2}\partial_{\mu_2}\psi_L(x_2)\partial_{\mu_3}\bar{\psi}_L(x_3)\gamma^{\nu_3}
\psi_L(x_3)\rangle_0^{(PV,\,c\,)},}\\[12pt]

{{\cal T}_4=3\int dx_1\,dx_2\,dx_3\,H^{\mu_1}_{\nu_1}(x_1)\,H^{\mu_2}_{\nu_2}(x_2)H^{\mu_3}_{\nu_3}(x_3)}\\[4pt]
{
\phantom{{\cal T}_1=3\int dx_1\,dx_2\,dx_3\,,H}
\langle\partial_{\mu_1}\bar{\psi}_L(x_1)\gamma^{\nu_1}\psi_L(x_1)\partial_{\mu_2}\bar{\psi}_L(x_2)\gamma^{\nu_2}\psi_L(x_2)\bar{\psi}_L(x_3)\gamma^{\nu_3}
\partial_{\mu_3}\psi_L(x_3)\rangle_0^{(PV,\,c\,)},}\\[12pt]
\end{array}
\label{triangledefs}
\end{equation}

An elementary application of Wick`s theorem and the properties of the traces of Dirac matrices -see (\ref{DiracM})-- leads to
\begin{equation}
\begin{array}{l}
{{\cal T}_1={\cal T}_1^{(even)}+{\cal T}_1^{(odd)},}\\[8pt]
{{\cal T}_1^{(even)}= -\int\prod\limits_{i=1}^{3}\dpi(2\pi)^4\,\delta(p_1+p_2+p_3)\,H^{\mu_1\nu_1}(p_1)\,H^{\mu_2\nu_2}(p_2)H^{\mu_3\nu_3}(p_3)\Big\{
\tr[\gamma_{\nu_1}\gamma_{\sigma_1}\gamma_{\nu_2}\gamma_{\sigma_2}\gamma_{\nu_3}\gamma_{\sigma_3}]}\\[4pt]
{\quad\idq\Big[ q_{\mu_1}q^{\sigma_1}(q+p_2)_{\mu_2}(q+p_2)^{\sigma_2}(q+p_2+p_3)_{\mu_3}(q+p_2+p_3)^{\sigma_3}}\\[8pt]
{\phantom{\quad\idq\Big[ q_{\mu_1}q^{\sigma_1}(q+p_2)_{\mu_2}(q+p_2)}
G^{(reg)}(q)\,G^{(reg)}(q+p_2)\,G^{(reg)}(q+p_2+p_3)\Big]
\Big\},}\\[12pt]
{{\cal T}_1^{(odd)}= -\int\prod\limits_{i=1}^{3}\dpi(2\pi)^4\,\delta(p_1+p_2+p_3)\,H^{\mu_1\nu_1}_{1}(p_1)\,H^{\mu_2\nu_2}_{1}(p_2)H^{\mu_3\nu_3}_{1}(p_3)\Big\{
\tr[\gamma_{\nu_1}\gamma_{\sigma_1}\gamma_{\nu_2}\gamma_{\sigma_2}\gamma_{\nu_3}\gamma_{\sigma_3}\gamma_5]}\\[4pt]
{\quad\idq\Big[ q_{\mu_1}q^{\sigma_1}(q+p_2)_{\mu_2}(q+p_2)^{\sigma_2}(q+p_2+p_3)_{\mu_3}(q+p_2+p_3)^{\sigma_3}}\\[8pt]
{\phantom{\quad\idq\Big[ q_{\mu_1}q^{\sigma_1}(q+p_2)_{\mu_2}(q+p_2)}
G^{(reg)}(q)\,G^{(reg)}(q+p_2)\,G^{(reg)}(q+p_2+p_3)\Big]
\Big\},}\\[12pt]

{{\cal T}_2={\cal T}_2^{(even)}+{\cal T}_2^{(odd)}}\\[8pt]
{{\cal T}_2^{(even)}={\cal T}_1^{(even)},\quad {\cal T}_2^{(odd)}=-{\cal T}_1^{(odd)}}\\[20pt]

{{\cal T}_3={\cal T}_3^{(even)}+{\cal T}_3^{(odd)}}\\[8pt]
{{\cal T}_3^{(even)}= -3\int\prod\limits_{i=1}^{3}\dpi(2\pi)^4\,\delta(p_1+p_2+p_3)\,H^{\mu_1\nu_1}(p_1)\,H^{\mu_2\nu_2}(p_2)H^{\mu_3\nu_3}(p_3)\Big\{
\tr[\gamma_{\nu_1}\gamma_{\sigma_1}\gamma_{\nu_2}\gamma_{\sigma_2}\gamma_{\nu_3}\gamma_{\sigma_3}]}\\[4pt]
{\quad\idq\Big[
q_{\mu_1}q^{\sigma_1}(q+p_2)_{\mu_2}(q+p_2)_{\mu_3}(q+p_2+p_3)^{\sigma_2}(q+p_2+p_3)^{\sigma_3}]}\\[4pt]
{\phantom{\quad\idq\Big[ q_{\mu_1}q^{\sigma_1}(q+p_2)_{\mu_2}(q+p_2)}
G^{(reg)}(q)\,G^{(reg)}(q+p_2)\,G^{(reg)}(q+p_2+p_3)\Big]
\Big\},}\\[12pt]
{{\cal T}_3^{(odd)}= -3\int\prod\limits_{i=1}^{3}\dpi(2\pi)^4\,\delta(p_1+p_2+p_3)\,H^{\mu_1\nu_1}(p_1)\,H^{\mu_2\nu_2}(p_2)H^{\mu_3\nu_3}(p_3)\Big\{
\tr[\gamma_{\nu_1}\gamma_{\sigma_1}\gamma_{\nu_2}\gamma_{\sigma_2}\gamma_{\nu_3}\gamma_{\sigma_3}\gamma_5]}\\[4pt]
{\quad\idq\Big[
q_{\mu_1}q^{\sigma_1}(q+p_2)_{\mu_2}(q+p_2)_{\mu_3}(q+p_2+p_3)^{\sigma_2}(q+p_2+p_3)^{\sigma_3}}\\[4pt]
{\phantom{\quad\idq\Big[ q_{\mu_1}q^{\sigma_1}(q+p_2)_{\mu_2}(q+p_2)}
G^{(reg)}(q)\,G^{(reg)}(q+p_2)\,G^{(reg)}(q+p_2+p_3)\Big]
\Big\},}\\[20pt]

{{\cal T}_4={\cal T}_4^{(even)}+{\cal T}_4^{(odd)}}\\[8pt]
{{\cal T}_4^{(even)}={\cal T}_3^{(even)},\quad {\cal T}_4^{(odd)}=-{\cal T}_3^{(odd)}.}
\end{array}
\label{trianglesvalues}
\end{equation}

Taking into account (\ref{triangle}) and (\ref{trianglesvalues}), one concludes that
\begin{equation*}
\langle (S_{kin})^3\rangle_0^{(PV,\,c\,)}=-\cfrac{i}{4}\big({\cal T}_1^{(even)}+{\cal T}_3^{(even)}\big)
\end{equation*}
We thus have shown that $\langle (S_{kin})^3\rangle_0^{(PV,\,c\,)}$ is parity-even.

\subsection{No parity-odd contribution from $\langle S_{kin}S_{spin}\rangle_0^{(PV,\,c\,)}$.}

From the definitions in (\ref{Sintexpan}) and (\ref{Hexpansion}), one gets
\begin{equation*}
\langle S_{kin}S_{spin}\rangle_0^{(PV,\,c\,)} =\cfrac{i}{2}\,\big({\mathbb B}_1+{\mathbb B}_2\big),
\end{equation*}
where
\begin{equation*}
\begin{array}{l}
{{\mathbb B}_1=\int dx_1\,dx_2\,H^{\mu_1}_{\nu_1}(x_1)\,H_{\mu_2\nu_2\rho_2}(x_2)\,\epsilon^{\mu_2\nu_2\rho_2\lambda}
\langle\bar{\psi}_L(x_1)\gamma^{\nu_1}\partial_{\mu_1}\psi_L(x_1)\bar{\psi}_L(x_2)\gamma_{\lambda}\psi_L(x_2)\rangle_0^{(PV,\,c\,)},}\\[12pt]
{{\mathbb B}_2=-\int dx_1\,dx_2\,H^{\mu_1}_{\nu_1}(x_1)\,H_{\mu_2\nu_2\rho_2}(x_2)\,\epsilon^{\mu_2\nu_2\rho_2\lambda}
\langle\partial_{\mu_1}\bar{\psi}_L(x_1)\gamma^{\nu_1}\psi_L(x_1)\bar{\psi}_L(x_2)\gamma_{\lambda}\psi_L(x_2)\rangle_0^{(PV,\,c\,)}.}
\end{array}
\end{equation*}

It turns out that
\begin{equation*}
\begin{array}{l}
{{\mathbb B}_1={\mathbb B}_1^{(even)}+{\mathbb B}_1^{(odd)}}\\[8pt]
{{\mathbb B}_1^{(even)}=-\frac{i}{2}\int\prod\limits_{i=1}^{2}\dpi(2\pi)^4\,\delta(p_1+p_2)\,H^{\mu_1\nu_1}(p_1)\,H_{\mu_2\nu_2\rho_2}(p_2)\Big\{
\epsilon^{\mu_2\nu_2\rho_2\lambda}\,\tr[\gamma_{\nu_1}\gamma_{\sigma_1}\gamma_{\lambda}\gamma_{\sigma_2}\gamma_5]}\\[4pt]
{\quad\quad\quad\quad\quad\quad\quad\quad\quad\quad\quad\quad\quad\idq\,
q_{\mu_1}q^{\sigma_1}(q+p_2)^{\sigma_2}\,
 G^{(reg)}(q)\,G^{(reg)}(q+p_2)\Big\},}\\[12pt]

{{\mathbb B}_1^{(odd)}=-\frac{i}{2}\int\prod\limits_{i=1}^{2}\dpi(2\pi)^4\,\delta(p_1+p_2)\,H^{\mu_1\nu_1}(p_1)\,H_{\mu_2\nu_2\rho_2}(p_2)\Big\{
\epsilon^{\mu_2\nu_2\rho_2\lambda}\,\tr[\gamma_{\nu_1}\gamma_{\sigma_1}\gamma_{\lambda}\gamma_{\sigma_2}]}\\[4pt]
{\quad\quad\quad\quad\quad\quad\quad\quad\quad\quad\quad\quad\quad\idq
\,q_{\mu_1}q^{\sigma_1}(q+p_2)^{\sigma_2}\,
 G^{(reg)}(q)\,G^{(reg)}(q+p_2)\Big\},}\\[12pt]

{{\mathbb B}_2={\mathbb B}_2^{(even)}+{\mathbb B}_2^{(odd)},}\\[8pt]
{{\mathbb B}_2^{(even)}={\mathbb B}_1^{(even)},\quad{\mathbb B}_2^{(odd)}=-{\mathbb B}_1^{(odd)}.}\\[20pt]
\end{array}
\end{equation*}
Hence,
\begin{equation*}
\langle S_{kin}S_{spin}\rangle_0^{(PV,\,c\,)} =i\,{\mathbb B}_1^{(even)}.
\label{bubblespinresult}
\end{equation*}
Thus, we have shown that $\langle S_{kin}S_{spin}\rangle_0^{(PV,\,c\,)}$ is parity-even.

\subsection{No parity-odd contribution from $\langle S_{kin}^2 S_{spin}\rangle_0^{(PV,\,c\,)}$.}

By taking into account (\ref{Sintexpan}}), (\ref{Hexpansion}) and (\ref{freeprop}), one gets

\begin{equation}
\langle S_{kin}^2 S_{spin}\rangle_0^{(PV,\,c\,)}= -\frac{1}{4}\,({\mathbb C}_1+{\mathbb C}_2+{\mathbb C}_3+{\mathbb C}_4),
\label{Skin2Sspin}
\end{equation}
where
\begin{equation}
\begin{array}{l}
{{\mathbb C}_1=\int dx_1\,dx_2\,dx_3\,H_{\mu\nu\rho}(x_1)\,H^{\mu_2}_{\nu_2}(x_2)H^{\mu_3}_{\nu_3}(x_3)}\\[4pt]
{
\phantom{{\cal T}_1=\int dx_1\,dx_2\,dx_3\,,H^{\mu_1}}\epsilon^{\mu\nu\rho\lambda}
\langle\bar{\psi}_L(x_1)\gamma_{\lambda}\psi_L(x_1)\bar{\psi}_L(x_2)\gamma^{\nu_2}\partial_{\mu_2}\psi_L(x_2)\bar{\psi}_L(x_3)\gamma^{\nu_3}
\partial_{\mu_3}\psi_L(x_3)\rangle_0^{(PV,\,c\,)},}\\[12pt]
{{\mathbb C}_2=\int dx_1\,dx_2\,dx_3\,H_{\mu\nu\rho}(x_1)\,H^{\mu_2}_{\nu_2}(x_2)H^{\mu_3}_{\nu_3}(x_3)}\\[4pt]
{
\phantom{{\cal T}_1=\int dx_1\,dx_2\,dx_3\,,H^{\mu_1}}\epsilon^{\mu\nu\rho\lambda}
\langle\bar{\psi}_L(x_1)\gamma_{\lambda}\psi_L(x_1)\partial_{\mu_2}\bar{\psi}_L(x_2)\gamma^{\nu_2}\psi_L(x_2)\partial_{\mu_3}\bar{\psi}_L(x_3)\gamma^{\nu_3}
\psi_L(x_3)\rangle_0^{(PV,\,c\,)},}\\[12pt]

{{\mathbb C}_3=-\int dx_1\,dx_2\,dx_3\,H_{\mu\nu\rho}(x_1)\,H^{\mu_2}_{\nu_2}(x_2)H^{\mu_3}_{\nu_3}(x_3)}\\[4pt]
{
\phantom{{\cal T}_1=\int dx_1\,dx_2\,dx_3\,,H^{\mu_1}}\epsilon^{\mu\nu\rho\lambda}
\langle\bar{\psi}_L(x_1)\gamma_{\lambda}\psi_L(x_1)\bar{\psi}_L(x_2)\gamma^{\nu_2}\partial_{\mu_2}\psi_L(x_2)\partial_{\mu_3}\bar{\psi}_L(x_3)\gamma^{\nu_3}
\psi_L(x_3)\rangle_0^{(PV,\,c\,)},}\\[12pt]

{{\mathbb C}_4=-\int dx_1\,dx_2\,dx_3\,H_{\mu\nu\rho}(x_1)\,H^{\mu_2}_{\nu_2}(x_2)H^{\mu_3}_{\nu_3}(x_3)}\\[4pt]
{
\phantom{{\cal T}_1=\int dx_1\,dx_2\,dx_3\,,H^{\mu_1}}\epsilon^{\mu\nu\rho}
\langle\bar{\psi}_L(x_1)\gamma_{\lambda}\psi_L(x_1)\partial_{\mu_2}\bar{\psi}_L(x_2)\gamma^{\nu_2}\psi_L(x_2)\bar{\psi}_L(x_3)\gamma^{\nu_3}
\partial_{\mu_3}\psi_L(x_3)\rangle_0^{(PV,\,c\,)}.}
\end{array}
\label{CesSkin2Spin}
\end{equation}

It can be shown that
\begin{equation}
\begin{array}{l}
{{\mathbb  C}_1={\mathbb C}_1^{(even)}+{\mathbb C}_1^{(odd)},}\\[8pt]
{{ \mathbb C}_1^{(even)}=i\int\prod\limits_{i=1}^{3}\dpi(2\pi)^4\,\delta(p_1+p_2+p_3)\,H_{\mu\nu\rho}(p_1)\,H^{\mu_2\nu_2}(p_2)H^{\mu_3\nu_3}(p_3)\Big\{}\\[4pt]
{
\epsilon^{\mu\nu\rho\lambda}\tr[\gamma_{\lambda}\gamma_{\sigma_1}\gamma_{\nu_2}\gamma_{\sigma_2}\gamma_{\nu_3}\gamma_{\sigma_3}\gamma_5]
\idq\Big[
q^{\sigma_1}(q+p_2)_{\mu_2}(q+p_2)^{\sigma_2}(q+p_2+p_3)_{\mu_3}(q+p_2+p_3)^{\sigma_3}\,}\\[8pt]
{\phantom{\idq\Big[
q^{\sigma_1}(q+p_2)_{\mu_2}(q+p_2)^{\sigma_2}(q+p_2+p_3)_{\mu_3}}
 G^{(reg)}(q)\,G^{(reg)}(q+p_2)\,G^{(reg)}(q+p_2+p_3)\Big]\Big\},}\\[12pt]

{{ \mathbb C}_1^{(odd)}=i\int\prod\limits_{i=1}^{3}\dpi(2\pi)^4\,\delta(p_1+p_2+p_3)\,H_{\mu\nu\rho}(p_1)\,H^{\mu_2\nu_2}(p_2)H^{\mu_3\nu_3}(p_3)\Big\{}\\
{
\epsilon^{\mu\nu\rho\lambda}\tr[\gamma_{\lambda}\gamma_{\sigma_1}\gamma_{\nu_2}\gamma_{\sigma_2}\gamma_{\nu_3}\gamma_{\sigma_3}]
\idq\Big[
q^{\sigma_1}(q+p_2)_{\mu_2}(q+p_2)^{\sigma_2}(q+p_2+p_3)_{\mu_3}(q+p_2+p_3)^{\sigma_3}\,}\\[8pt]
{\phantom{\idq\Big[
q^{\sigma_1}(q+p_2)_{\mu_2}(q+p_2)^{\sigma_2}(q+p_2+p_3)_{\mu_3}}
 G^{(reg)}(q)\,G^{(reg)}(q+p_2)\,G^{(reg)}(q+p_2+p_3)\Big]\Big\},}\\[20pt]

{{ \mathbb C}_2={ \mathbb C}_2^{(even)}+{ \mathbb C}_2^{(odd)},}\\[8pt]
{{ \mathbb C}_2^{(even)}={ \mathbb C}_1^{(even)},\quad { \mathbb C}_2^{(odd)}=-{ \mathbb C}_1^{(odd)}}
\end{array}
\label{triangle12Ces}
\end{equation}
and

\begin{equation}
\begin{array}{l}

{{ \mathbb C}_3={ \mathbb C}_{31}^{(even)}+{ \mathbb C}_{31}^{(odd)}+{ \mathbb C}_{32}^{(even)}+{ \mathbb C}_{32}^{(odd)},}\\[8pt]

{{\mathbb C}_{31}^{(even)}=\frac{i}{2}\int\prod\limits_{i=1}^{3}\dpi(2\pi)^4\,\delta(p_1+p_2+p_3)\,H_{\mu\nu\rho}(p_1)\,H^{\mu_2\nu_2}(p_2)H^{\mu_3\nu_3}(p_3)\Big\{}\\
{\quad\quad\quad\quad\quad\quad\quad\quad
\epsilon^{\mu\nu\rho\lambda}\tr[\gamma_{\lambda}\gamma_{\sigma_1}\gamma_{\nu_2}\gamma_{\sigma_2}\gamma_{\nu_3}\gamma_{\sigma_3}\gamma_5]}\\[4pt]
{\idq\Big[ q^{\sigma_1}(q+p_2)_{\mu_2}(q+p_2)_{\mu_3}(q+p_2)^{\sigma_2}(q+p_2+p_3)^{\sigma_3}}\\[8pt]
{\phantom{\idq\Big[ q^{\sigma_1}(q+p_2)_{\mu_2}(q+p_2)_{\mu_3}(q+p_2)^{\sigma_2}}
 G^{(reg)}(q)\,G^{(reg)}(q+p_2)\,G^{(reg)}(q+p_2+p_3)\Big]
\Big\},}\\[12pt]

{{\mathbb C}_{31}^{(odd)}=\frac{i}{2}\int\prod\limits_{i=1}^{3}\dpi(2\pi)^4\,\delta(p_1+p_2+p_3)\,H_{\mu\nu\rho}(p_1)\,H^{\mu_2\nu_2}(p_2)H^{\mu_3\nu_3}(p_3)\Big\{}\\
{\quad\quad\quad\quad\quad\quad\quad\quad
\epsilon^{\mu\nu\rho\lambda}\tr[\gamma_{\lambda}\gamma_{\sigma_1}\gamma_{\nu_2}\gamma_{\sigma_2}\gamma_{\nu_3}\gamma_{\sigma_3}]}\\[4pt]
{\idq\Big[ q^{\sigma_1}(q+p_2)_{\mu_2}(q+p_2)_{\mu_3}(q+p_2)^{\sigma_2}(q+p_2+p_3)^{\sigma_3}}\\[8pt]
{\phantom{\idq\Big[ q^{\sigma_1}(q+p_2)_{\mu_2}(q+p_2)_{\mu_3}(q+p_2)^{\sigma_2}}
 G^{(reg)}(q)\,G^{(reg)}(q+p_2)\,G^{(reg)}(q+p_2+p_3)\Big]
\Big\},}\\[20pt]

{{\mathbb C}_{32}^{(even)}=-\frac{i}{2}\int\prod\limits_{i=1}^{3}\dpi(2\pi)^4\,\delta(p_1+p_2+p_3)\,H_{\mu\nu\rho}(p_1)\,H^{\mu_2\nu_2}(p_2)H^{\mu_3\nu_3}(p_3)\Big\{}\\
{\quad\quad\quad\quad\quad\quad\quad\quad
\epsilon^{\mu\nu\rho\lambda}\tr[\gamma_{\lambda}\gamma_{\sigma_3}\gamma_{\nu_3}\gamma_{\sigma_2}\gamma_{\nu_2}\gamma_{\sigma_1}\gamma_5]}\\[4pt]
{\idq\Big[
q_{\mu_2}q^{\sigma_1}(q+p_2)^{\sigma_2}(q+p_2+p_3)_{\mu_3}(q+p_2+p_3)^{\sigma_3}}\\[8pt]
{\phantom{\idq\Big[
q_{\mu_2}q^{\sigma_1}(q+p_2)^{\sigma_2}(q+p_2+p_3)_{\mu_3}}
G^{(reg)}(q)\,G^{(reg)}(q+p_2)\,G^{(reg)}(q+p_2+p_3)\Big]
\Big\},}\\[12pt]

{{\mathbb C}_{32}^{(odd)}=-\frac{i}{2}\int\prod\limits_{i=1}^{3}\dpi(2\pi)^4\,\delta(p_1+p_2+p_3)\,H_{\mu\nu\rho}(p_1)\,H^{\mu_2\nu_2}(p_2)H^{\mu_3\nu_3}(p_3)\Big\{}\\
{\quad\quad\quad\quad\quad\quad\quad\quad
\epsilon^{\mu\nu\rho\lambda}\tr[\gamma_{\lambda}\gamma_{\sigma_3}\gamma_{\nu_3}\gamma_{\sigma_2}\gamma_{\nu_2}\gamma_{\sigma_1}]}\\[4pt]
{\idq\Big[
q_{\mu_2}q^{\sigma_1}(q+p_2)^{\sigma_2}(q+p_2+p_3)_{\mu_3}(q+p_2+p_3)^{\sigma_3}}\\[8pt]
{\phantom{\idq\Big[
q_{\mu_2}q^{\sigma_1}(q+p_2)^{\sigma_2}(q+p_2+p_3)_{\mu_3}}
G^{(reg)}(q)\,G^{(reg)}(q+p_2)\,G^{(reg)}(q+p_2+p_3)\Big]
\Big\},}\\[20pt]

{{ \mathbb C}_4={ \mathbb C}_{41}^{(even)}+{ \mathbb C}_{41}^{(odd)}+{ \mathbb C}_{42}^{(even)}+{ \mathbb C}_{42}^{(odd)},}\\[8pt]
{{ \mathbb C}_{41}^{(even)}={ \mathbb C}_{31}^{(even)},\quad { \mathbb C}_{41}^{(odd)}=-{ \mathbb C}_{31}^{(odd)},\quad
{ \mathbb C}_{42}^{(even)}={ \mathbb C}_{32}^{(even)},\quad { \mathbb C}_{42}^{(odd)}=-{ \mathbb C}_{32}^{(odd)}.
}
\end{array}
\label{triangle34Ces}
\end{equation}

Equations (\ref{Skin2Sspin}), (\ref{CesSkin2Spin}), (\ref{triangle12Ces}) and (\ref{triangle34Ces}) lead to the conclusion that
\begin{equation}
\langle S_{kin}^2 S_{spin}\rangle_0^{(PV,\,c\,)}= -\frac{1}{2}\,({\mathbb C}_1^{(even)}+{\mathbb C}_{31}^{(even)}+{\mathbb C}_{32}^{(even)}).
\label{Skin2Sspinfinal}
\end{equation}
Again, no odd parity contribution arises here.

\newpage

\subsection{No parity-odd contribution from $\langle S_{spin}^2\rangle_0^{(PV,\,c\,)}$.}

The definitions in (\ref{Sintexpan})  lead to
\begin{equation*}
\begin{array}{l}
{\langle S_{spin}^2\rangle_0^{(PV,\,c\,)} =}\\[8pt]
{\int dx_1\,dx_2\,H_{\mu_1\nu_1\rho_1}(x_1)\,H_{\mu_2\nu_2\rho_2}(x_2)\,\epsilon^{\mu_1\nu_1\rho_1\lambda_1}\epsilon^{\mu_2\nu_2\rho_2\lambda_2}
\langle\bar{\psi}(x_1)\gamma_{\lambda_1}P_L\psi(x_1)\bar{\psi}(x_2)\gamma_{\lambda_2}P_L\psi(x_2)\rangle_0^{(PV,\,c\,)}=}\\[8pt]
{\int\prod\limits_{i=1}^{2}\dpi(2\pi)^4\,\delta(p_1+p_2)\,H_{\mu_1\nu_1\rho_1}(p_1)H_{\mu_2\nu_2\rho_2}(p_2)\Big\{
\epsilon^{\mu_1\nu_1\rho_1\lambda_1}\epsilon^{\mu_2\nu_2\rho_2\lambda_2}
\tr[\gamma_{\lambda_1}\gamma_{\sigma_1}\gamma_{\lambda_2}\gamma_{\sigma_2}P_R]}\\[4pt]
{\quad\quad\quad\quad\quad\quad\quad\quad\quad\idq q^{\sigma_1}(q+p_2)^{\sigma_2}\,G^{(reg)}(q)\,G^{(reg)}(q+p_2)\Big\}.
}
\end{array}
\end{equation*}
Next, taking into account that
\begin{equation*}
\begin{array}{l}
{\tr[\gamma_{\lambda_1}\gamma_{\sigma_1}\gamma_{\lambda_2}\gamma_{\sigma_2}\gamma_5]\, q^{\sigma_2}q^{\sigma_2}=0,\quad\text{and that}}\\[8pt]
{\idq q^{\sigma_1}\,G^{(reg)}(q)\,G^{(reg)}(q+p_2)=p_2^{\sigma_1}\,f(p_2^2,\Lambda^2),}
\end{array}
\end{equation*}
one easily shows that the odd parity contribution to $\langle S_{spin}^2\rangle_0^{(PV,\,c\,)}$ vanishes.

\newpage
\subsection{No parity-odd contribution from $\langle S_{kin}^4\rangle_0^{(PV,\,c\,)}$.}

Let us finally show that there is a total cancellation of the partial parity-odd contributions which arise in the computation of $\langle S_{kin}^4\rangle_0^{(PV,\,c\,)}$. It is plain that we have
\begin{equation}
\langle (S_{kin})^4\rangle_0^{(PV,\,c\,)}=\cfrac{1}{16}\,\big({\cal B}_1+{\cal B}_2+{\cal B}_3+{\cal B}_4+{\cal B}_5+{\cal B}_6\big),
\label{boxes}
\end{equation}
where
\begin{equation}
\begin{array}{l}
{{\cal B}_1=\int dx_1\,dx_2\,dx_3\,dx_4\,\Big\{H^{\mu_1}_{\nu_1}(x_1)\,H^{\mu_2}_{\nu_2}(x_2)H^{\mu_3}_{\nu_3}(x_3)H^{\mu_4}_{\nu_4}(x_4)}\\[4pt]
{\quad
\langle\bar{\psi}_L(x_1)\gamma^{\nu_1}\partial_{\mu_1}\psi_L(x_1)\bar{\psi}_L(x_2)\gamma^{\nu_2}\partial_{\mu_2}\psi_L(x_2)\bar{\psi}_L(x_3)\gamma^{\nu_3}
\partial_{\mu_3}\psi_L(x_3)\bar{\psi}_L(x_4)\gamma^{\nu_4}
\partial_{\mu_4}\psi_L(x_4)\rangle_0^{(PV,\,c\,)}\Big\},}\\[20pt]
{{\cal B}_2=\int dx_1\,dx_2\,dx_3\,dx_4\,\Big\{H^{\mu_1}_{\nu_1}(x_1)\,H^{\mu_2}_{\nu_2}(x_2)H^{\mu_3}_{\nu_3}(x_3)H^{\mu_4}_{\nu_4}(x_4)}\\[4pt]
{\quad
\langle\partial_{\mu_1}\bar{\psi}_L(x_1)\gamma^{\nu_1}\psi_L(x_1)\partial_{\mu_2}\bar{\psi}_L(x_2)\gamma^{\nu_2}\psi_L(x_2)\partial_{\mu_3}\bar{\psi}_L(x_3)
\gamma^{\nu_3}\psi_L(x_3)\partial_{\mu_4}\bar{\psi}_L(x_4)\gamma^{\nu_4}
\psi_L(x_4)\rangle_0^{(PV,\,c\,)}\Big\},}\\[20pt]

{{\cal B}_3=-4\int dx_1\,dx_2\,dx_3\,dx_4\,\Big\{H^{\mu_1}_{\nu_1}(x_1)\,H^{\mu_2}_{\nu_2}(x_2)H^{\mu_3}_{\nu_3}(x_3)H^{\mu_4}_{\nu_4}(x_4)}\\[4pt]
{\quad
\langle\bar{\psi}_L(x_1)\gamma^{\nu_1}\partial_{\mu_1}\psi_L(x_1)\partial_{\mu_2}\bar{\psi}_L(x_2)\gamma^{\nu_2}\psi_L(x_2)\partial_{\mu_3}\bar{\psi}_L(x_3)
\gamma^{\nu_3}\psi_L(x_3)\partial_{\mu_4}\bar{\psi}_L(x_4)\gamma^{\nu_4}
\psi_L(x_4)\rangle_0^{(PV,\,c\,)}\Big\},}\\[20pt]

{{\cal B}_4=-4\int dx_1\,dx_2\,dx_3\,dx_4\,\Big\{H^{\mu_1}_{\nu_1}(x_1)\,H^{\mu_2}_{\nu_2}(x_2)H^{\mu_3}_{\nu_3}(x_3)H^{\mu_4}_{\nu_4}(x_4)}\\[4pt]
{\quad
\langle\partial_{\mu_1}\bar{\psi}_L(x_1)\partial_{\mu_1}\psi_L(x_1)\bar{\psi}_L(x_2)\gamma^{\nu_2}\partial_{\mu_2}\psi_L(x_2)\bar{\psi}_L(x_3)\gamma^{\nu_3}
\partial_{\mu_3}\psi_L(x_3)\bar{\psi}_L(x_4)\gamma^{\nu_4}
\partial_{\mu_4}\psi_L(x_4)\rangle_0^{(PV,\,c\,)}\Big\},}\\[20pt]

{{\cal B}_5=3\int dx_1\,dx_2\,dx_3\,dx_4\,\Big\{H^{\mu_1}_{\nu_1}(x_1)\,H^{\mu_2}_{\nu_2}(x_2)H^{\mu_3}_{\nu_3}(x_3)H^{\mu_4}_{\nu_4}(x_4)}\\[4pt]
{\quad
\langle\partial_{\mu_1}\bar{\psi}_L(x_1)\psi_L(x_1)\partial_{\mu_2}\bar{\psi}_L(x_2)\gamma^{\nu_2}\psi_L(x_2)\bar{\psi}_L(x_3)\gamma^{\nu_3}
\partial_{\mu_3}\psi_L(x_3)\bar{\psi}_L(x_4)\gamma^{\nu_4}
\partial_{\mu_4}\psi_L(x_4)\rangle_0^{(PV,\,c\,)}\Big\},}\\[20pt]

{{\cal B}_6=3\int dx_1\,dx_2\,dx_3\,dx_4\,\Big\{H^{\mu_1}_{\nu_1}(x_1)\,H^{\mu_2}_{\nu_2}(x_2)H^{\mu_3}_{\nu_3}(x_3)H^{\mu_4}_{\nu_4}(x_4)}\\[4pt]
{\quad
\langle\bar{\psi}_L(x_1)\gamma^{\nu_1}\partial_{\mu_1}\psi_L(x_1)\bar{\psi}_L(x_2)\gamma^{\nu_2}\partial_{\mu_2}\psi_L(x_2)\partial_{\mu_3}\bar{\psi}_L(x_3)
\gamma^{\nu_3}\psi_L(x_3)\partial_{\mu_4}\bar{\psi}_L(x_4)\gamma^{\nu_4}
\psi_L(x_4)\rangle_0^{(PV,\,c\,)}\Big\}.}
\end{array}
\label{boxdefs}
\end{equation}

Then, a straightforward application of Wick's theorem leads to the following results
\begin{equation}
\begin{array}{l}
{{\cal B}_1={\cal B}_1^{(even)}+{\cal B}_1^{(odd)},}\\[8pt]
{{\cal B}_1^{(even)}= -3\int\prod\limits_{i=1}^{4}\dpi(2\pi)^4\,\delta(p_1+p_2+p_3+p_4)\,H^{\mu_1\nu_1}(p_1)\,H^{\mu_2\nu_2}(p_2)H^{\mu_3\nu_3}(p_3)H^{\mu_4\nu_4}(p_4)\Big\{}\\
{\quad\quad\quad\quad\quad\quad\quad\quad
\tr[\gamma_{\nu_1}\gamma_{\sigma_1}\gamma_{\nu_2}\gamma_{\sigma_2}\gamma_{\nu_3}\gamma_{\sigma_3}\gamma_{\nu_4}\gamma_{\sigma_4}]}\\[4pt]
{\idq\Big[q_{\mu_1}q^{\sigma_1}(q+p_2)_{\mu_2}(q+p_2)^{\sigma_2}(q+p_2+p_3)_{\mu_3}(q+p_2+p_3)^{\sigma_3}(q-p_1)_{\mu_4}(q-p_1)^{\sigma_4}}\\[4pt]
{\phantom{\idq\Big[q}
G^{(reg)}(q)\,G^{(reg)}(q+p_2)\,G^{(reg)}(q+p_2+p_3)\,G^{(reg)}(q+p_2+p_3+p_4)\Big]\Big\},}\\[12pt]
{{\cal B}_1^{(odd)}= -3\int\prod\limits_{i=1}^{4}\dpi(2\pi)^4\,\delta(p_1+p_2+p_3+p_4)\,H^{\mu_1\nu_1}(p_1)\,H^{\mu_2\nu_2}(p_2)H^{\mu_3\nu_3}(p_3)
H^{\mu_4\nu_4}(p_4)\Big\{}\\
{\quad\quad\quad\quad\quad\quad\quad\quad
\tr[\gamma_{\nu_1}\gamma_{\sigma_1}\gamma_{\nu_2}\gamma_{\sigma_2}\gamma_{\nu_3}\gamma_{\sigma_3}\gamma_{\nu_4}\gamma_{\sigma_4}\gamma_5]}\\[4pt]
{\idq\Big[q_{\mu_1}q^{\sigma_1}(q+p_2)_{\mu_2}(q+p_2)^{\sigma_2}(q+p_2+p_3)_{\mu_3}(q+p_2+p_3)^{\sigma_3}(q-p_1)_{\mu_4}(q-p_1)^{\sigma_4}}\\[4pt]
{\phantom{\idq\Big[q}
G^{(reg)}(q)\,G^{(reg)}(q+p_2)\,G^{(reg)}(q+p_2+p_3)\,G^{(reg)}(q+p_2+p_3+p_4)\Big]\Big\},}\\[20pt]

{{\cal B}_2={\cal B}_2^{(even)}+{\cal B}_2^{(odd)},}\\[8pt]
{{\cal B}_2^{(even)}={\cal B}_1^{(even)},\quad{\cal B}_2^{(odd)}=-{\cal B}_1^{(odd)},}\\[20pt]
\label{box12s}
\end{array}
\end{equation}

\begin{equation}
\begin{array}{l}
{{\cal B}_3={\cal B}_3^{(even)}+{\cal B}_3^{(odd)},}\\[8pt]
{{\cal B}_3^{(even)}= -12\int\prod\limits_{i=1}^{4}\dpi(2\pi)^4\,\delta(p_1+p_2+p_3+p_4)\,H^{\mu_1\nu_1}(p_1)\,H^{\mu_2\nu_2}(p_2)H^{\mu_3\nu_3}(p_3)H^{\mu_4\nu_4}(p_4)\Big\{}\\
{\quad\quad\quad\quad\quad\quad\quad\quad
\tr[\gamma_{\nu_1}\gamma_{\sigma_1}\gamma_{\nu_2}\gamma_{\sigma_2}\gamma_{\nu_3}\gamma_{\sigma_3}\gamma_{\nu_4}\gamma_{\sigma_4}]}\\[4pt]
{\idq\Big[q^{\sigma_1}(q+p_2)_{\mu_2}(q+p_2)^{\sigma_2}(q+p_2+p_3)_{\mu_3}(q+p_2+p_3)^{\sigma_3}(q-p_1)_{\mu_4}(q-p_1)_{\mu_1}(q-p_1)^{\sigma_4}}\\[4pt]
{\phantom{\idq\Big[q}
G^{(reg)}(q)\,G^{(reg)}(q+p_2)\,G^{(reg)}(q+p_2+p_3)\,G^{(reg)}(q+p_2+p_3+p_4)\Big]\Big\},}\\[12pt]
{{\cal B}_3^{(odd)}= -12\int\prod\limits_{i=1}^{4}\dpi(2\pi)^4\,\delta(p_1+p_2+p_3+p_4)\,H^{\mu_1\nu_1}(p_1)\,H^{\mu_2\nu_2}(p_2)H^{\mu_3\nu_3}(p_3)H^{\mu_4\nu_4}(p_4)\Big\{}\\
{\quad\quad\quad\quad\quad\quad\quad\quad
\tr[\gamma_{\nu_1}\gamma_{\sigma_1}\gamma_{\nu_2}\gamma_{\sigma_2}\gamma_{\nu_3}\gamma_{\sigma_3}\gamma_{\nu_4}\gamma_{\sigma_4}\gamma_5]}\\[4pt]
{\idq\Big[q^{\sigma_1}(q+p_2)_{\mu_2}(q+p_2)^{\sigma_2}(q+p_2+p_3)_{\mu_3}(q+p_2+p_3)^{\sigma_3}(q-p_1)_{\mu_4}(q-p_1)_{\mu_1}(q-p_1)^{\sigma_4}}\\[4pt]
{\phantom{\idq\Big[q}
G^{(reg)}(q)\,G^{(reg)}(q+p_2)\,G^{(reg)}(q+p_2+p_3)\,G^{(reg)}(q+p_2+p_3+p_4)\Big]\Big\},}\\[12pt]

{{\cal B}_4={\cal B}_4^{(even)}+{\cal B}_4^{(odd)},}\\[8pt]
{{\cal B}_4^{(even)}={\cal B}_3^{(even)},\quad{\cal B}_4^{(odd)}=-{\cal B}_3^{(odd)}.}
\label{box34s}
\end{array}
\end{equation}
and
\begin{equation}
\begin{array}{l}
{{\cal B}_5={\cal B}_{51}^{(even)}+{\cal B}_{51}^{(odd)}+{\cal B}_{52}^{(even)}+{\cal B}_{52}^{(odd)}+{\cal B}_{53}^{(even)}+{\cal B}_{53}^{(odd)},}\\[2pt]
{{\cal B}_{51}^{(even)}= -3\int\prod\limits_{i=1}^{4}\dpi(2\pi)^4\,\delta(p_1+p_2+p_3+p_4)\,H^{\mu_1\nu_1}(p_1)\,H^{\mu_2\nu_2}(p_2)H^{\mu_3\nu_3}(p_3)H^{\mu_4\nu_4}(p_4)\Big\{}\\[1pt]
{\quad\quad\quad\quad\quad\quad\quad\quad
\tr[\gamma_{\nu_1}\gamma_{\sigma_1}\gamma_{\nu_2}\gamma_{\sigma_2}\gamma_{\nu_3}\gamma_{\sigma_3}\gamma_{\nu_4}\gamma_{\sigma_4}]}\\[0.5pt]
{\idq\Big[q_{\mu_2}q^{\sigma_1}(q+p_2)^{\sigma_2}(q+p_2+p_3)_{\mu_3}(q+p_2+p_3)^{\sigma_3}(q-p_1)_{\mu_4}(q-p_1)_{\mu_1}(q-p_1)^{\sigma_4}}\\[1pt]
{\phantom{\idq\Big[q}
G^{(reg)}(q)\,G^{(reg)}(q+p_2)\,G^{(reg)}(q+p_2+p_3)\,G^{(reg)}(q+p_2+p_3+p_4)\Big]\Big\},}\\[20pt]
{{\cal B}_{51}^{(odd)}= 3\int\prod\limits_{i=1}^{4}\dpi(2\pi)^4\,\delta(p_1+p_2+p_3+p_4)\,H^{\mu_1\nu_1}(p_1)\,H^{\mu_2\nu_2}(p_2)H^{\mu_3\nu_3}(p_3)H^{\mu_4\nu_4}(p_4)\Big\{}\\[1pt]
{\quad\quad\quad\quad\quad\quad\quad\quad
\tr[\gamma_{\nu_1}\gamma_{\sigma_1}\gamma_{\nu_2}\gamma^{\sigma_2}\gamma_{\nu_3}\gamma_{\sigma_3}\gamma_{\nu_4}\gamma_{\sigma_4}\gamma_5]}\\[0.5pt]
{\idq\Big[q_{\mu_2}q^{\sigma_1}(q+p_2)_{\sigma_2}(q+p_2+p_3)_{\mu_3}(q+p_2+p_3)^{\sigma_3}(q-p_1)_{\mu_4}(q-p_1)_{\mu_1}(q-p_1)^{\sigma_4}}\\[1pt]
{\phantom{\idq\Big[q}
G^{(reg)}(q)\,G^{(reg)}(q+p_2)\,G^{(reg)}(q+p_2+p_3)\,G^{(reg)}(q+p_2+p_3+p_4)\Big]\Big\},}\\[20pt]

{{\cal B}_{52}^{(even)}= -3\int\prod\limits_{i=1}^{4}\dpi(2\pi)^4\,\delta(p_1+p_2+p_3+p_4)\,H^{\mu_1\nu_1}(p_1)\,H^{\mu_2\nu_2}(p_2)H^{\mu_3\nu_3}(p_3)H^{\mu_4\nu_4}(p_4)\Big\{}\\[1pt]
{\quad\quad\quad\quad\quad\quad\quad\quad
\tr[\gamma_{\nu_1}\gamma_{\sigma_1}\gamma_{\nu_2}\gamma_{\sigma_2}\gamma_{\nu_3}\gamma_{\sigma_3}\gamma_{\nu_4}\gamma_{\sigma_4}]}\\[0.5pt]
{\idq\Big[q_{\mu_1}q^{\sigma_1}(q+p_2)_{\mu_2}(q+p_2)_{\mu_3}(q+p_2)_{\sigma_2}(q+p_2+p_3)_{\mu_4}(q+p_2+p_3)^{\sigma_3}(q-p_1)^{\sigma_4}}\\[1pt]
{\phantom{\idq\Big[q}
G^{(reg)}(q)\,G^{(reg)}(q+p_2)\,G^{(reg)}(q+p_2+p_3)\,G^{(reg)}(q+p_2+p_3+p_4)\Big]\Big\},}\\[20pt]
{{\cal B}_{52}^{(odd)}= -3\int\prod\limits_{i=1}^{4}\dpi(2\pi)^4\,\delta(p_1+p_2+p_3+p_4)\,H^{\mu_1\nu_1}(p_1)\,H^{\mu_2\nu_2}(p_2)H^{\mu_3\nu_3}(p_3)H^{\mu_4\nu_4}(p_4)\Big\{}\\[1pt]
{\quad\quad\quad\quad\quad\quad\quad\quad
\tr[\gamma_{\nu_1}\gamma_{\sigma_1}\gamma_{\nu_2}\gamma_{\sigma_2}\gamma_{\nu_3}\gamma_{\sigma_3}\gamma_{\nu_4}\gamma_{\sigma_4}\gamma_5]}\\[0.5pt]
{\idq\Big[q_{\mu_1}q^{\sigma_1}(q+p_2)_{\mu_2}(q+p_2)_{\mu_3}(q+p_2)_{\sigma_2}(q+p_2+p_3)_{\mu_4}(q+p_2+p_3)^{\sigma_3}(q-p_1)^{\sigma_4}}\\[1pt]
{\phantom{\idq\Big[q}
G^{(reg)}(q)\,G^{(reg)}(q+p_2)\,G^{(reg)}(q+p_2+p_3)\,G^{(reg)}(q+p_2+p_3+p_4)\Big]\Big\},}\\[8pt]
\label{box512s}
\end{array}
\end{equation}

\begin{equation}
\begin{array}{l}
{{\cal B}_{53}^{(even)}= -3\int\prod\limits_{i=1}^{4}\dpi(2\pi)^4\,\delta(p_1+p_2+p_3+p_4)\,H^{\mu_1\nu_1}(p_1)\,H^{\mu_2\nu_2}(p_2)H^{\mu_3\nu_3}(p_3)H^{\mu_4\nu_4}(p_4)\Big\{}\\[1pt]
{\quad\quad\quad\quad\quad\quad\quad\quad
\tr[\gamma_{\nu_1}\gamma_{\sigma_1}\gamma_{\nu_2}\gamma_{\sigma_2}\gamma_{\nu_3}\gamma_{\sigma_3}\gamma_{\nu_4}\gamma_{\sigma_4}]}\\[0.5pt]
{\idq\Big[q^{\sigma_1}(q+p_2)_{\mu_2}(q+p_2)_{\mu_3}(q+p_2)^{\sigma_2}(q+p_2+p_3)^{\sigma_3}(q-p_1)_{\mu_4}(q-p_1)_{\mu_1}(q-p_1)^{\sigma_4}}\\[1pt]
{\phantom{\idq\Big[q}
G^{(reg)}(q)\,G^{(reg)}(q+p_2)\,G^{(reg)}(q+p_2+p_3)\,G^{(reg)}(q+p_2+p_3+p_4)\Big]\Big\},}\\[8pt]
{{\cal B}_{53}^{(odd)}= 3\int\prod\limits_{i=1}^{4}\dpi(2\pi)^4\,\delta(p_1+p_2+p_3+p_4)\,H^{\mu_1\nu_1}(p_1)\,H^{\mu_2\nu_2}(p_2)H^{\mu_3\nu_3}(p_3)H^{\mu_4\nu_4}(p_4)\Big\{}\\[1pt]
{\quad\quad\quad\quad\quad\quad\quad\quad
\tr[\gamma_{\nu_1}\gamma_{\sigma_1}\gamma_{\nu_2}\gamma^{\sigma_2}\gamma_{\nu_3}\gamma_{\sigma_3}\gamma_{\nu_4}\gamma_{\sigma_4}\gamma_5]}\\[0.5pt]
{\idq\Big[q_{\mu_2}q^{\sigma_1}(q+p_2)_{\sigma_2}(q+p_2+p_3)_{\mu_3}(q+p_2+p_3)^{\sigma_3}(q-p_1)_{\mu_4}(q-p_1)_{\mu_1}(q-p_1)^{\sigma_4}}\\[1pt]
{\phantom{\idq\Big[q}
G^{(reg)}(q)\,G^{(reg)}(q+p_2)\,G^{(reg)}(q+p_2+p_3)\,G^{(reg)}(q+p_2+p_3+p_4)\Big]\Big\},}
\label{box53s}
\end{array}
\end{equation}
and, finally,

\begin{equation}
\begin{array}{l}
{{\cal B}_6={\cal B}_{61}^{(even)}+{\cal B}_{61}^{(odd)}+{\cal B}_{62}^{(even)}+{\cal B}_{62}^{(odd)}+{\cal B}_{63}^{(even)}+{\cal B}_{63}^{(odd)},}\\[2pt]
{{\cal B}_{61}^{(even)}={\cal B}_{51}^{(even)},\quad {\cal B}_{61}^{(odd)}=-{\cal B}_{51}^{(odd)},\quad
{\cal B}_{62}^{(even)}={\cal B}_{52}^{(even)},}\\[4pt]
{{\cal B}_{62}^{(odd)}=-{\cal B}_{52}^{(odd)},\quad
{\cal B}_{63}^{(even)}={\cal B}_{53}^{(even)},\quad {\cal B}_{63}^{(odd)}=-{\cal B}_{53}^{(odd)}.
}
\label{box6s}
\end{array}
\end{equation}

Substituting (\ref{box12s}), (\ref{box34s}), (\ref{box512s}), (\ref{box53s}) and (\ref{box6s}) in (\ref{boxes}), one gets

\begin{equation*}
\langle (S_{kin})^4\rangle_0^{(PV,\,c\,)}=\cfrac{1}{8}\,\big({\cal B}_1^{(even)}+{\cal B}_3^{(even)}+{\cal B}_{51}^{(even)}+{\cal B}_{52}^{(even)}
+{\cal B}_{53}^{(even)}\big),
\end{equation*}
which is, obviously, free from parity-odd contributions.

In the previous subsections we have shown that no summand on the right hand side of (\ref{GPV}) carries a parity-odd contribution. Hence, as discussed in the paragraph below (\ref{GPV}), the one, two, three and four-point contribution to the regularized gravitational effective action ${\cal W}^{(PV)}[h_{\mu\nu}]$ are free from parity-odd terms.

Finally, the reader might wonder why no $(s-1)m$ occurs in the numerators of integrands of the final results displayed above.
The reason is that such contributions come hand in hand with the matrix $P_L\gamma^{\mu}P_L$, which vanishes. See appendix
for a discussion of this result.

\section{The renormalized functions in the BPHZL renormalization scheme.}

Let us first recall that the main purpose of this paper is to show that the renormalized gravitational effective action obtained from
 ${\cal G}^{(PV)}[h_{\mu\nu}]$ in (\ref{GPV}) carries no parity-odd contributions up to order four in the number of graviton fields.
This we shall do by using the BPHZL renormalization algorithm discussed in \cite{Clark:1976ym}. One word of caution before we move on, it is not the purpose of this paper to explain the BPHZL renormalization method in any detail, for discussing it properly will demand the inclusion in the paper of many pages and that material has been already covered in \cite{Lowenstein:1975ug}. Hence, we shall borrow from  \cite{Clark:1976ym}, and not  prove, the relevant formulae. 

Let
\begin{equation}
F[p,m(s-1),M_i,c_i,\varepsilon] = \idq\; I[q,p,m(s-1),M_i,c_i,\varepsilon]
\label{regint}
\end{equation}
generically denote any of the one-loop Feynman integrals --i.e., the integrals over $q$-- in section 4 (e.g., the integrals over $q$ in ${\cal B}_1$ in (\ref{box12s})). In the previous integral, $p$ stands for the external momenta.

 Let
\begin{equation}
 I[q,p,m(s-1),\varepsilon]
\label{unregint}
\end{equation}
be the unregularized  integrand associated to the Feynman integral in  (\ref{regint}), i.e., the function obtained by taking the limit $M_i\rightarrow \infty$ of $I[q,p,m(s-1),M_i,c_i,\varepsilon]$. Then, as it is discussed in \cite{Lowenstein:1975ps}, we associate to
$F[p,m(s-1), M_i,c_i,\varepsilon]$ above an ultraviolet (UV) degree, say $d$, and an infrared (IR) degree, say $r$, of divergence.
These degrees are obtained by working out the following asymptotic behaviours:
\begin{equation*}
\begin{array}{l}
{ I[\lambda q,\lambda p,m((\lambda s)-1),\varepsilon]\sim   \lambda^{d-4} \quad\text{as}\quad\lambda\rightarrow\infty,}\\[4pt]
{ I[\lambda q,\lambda p,m\lambda( s-1),\varepsilon]\sim   \lambda^{r-4} \quad\text{as}\quad\lambda\rightarrow 0.}
\end{array}
\end{equation*}

The next step in  the BPHZL subtraction procedure --see \cite{Lowenstein:1975ps} and \cite{Clark:1976ym}-- is to associate to the
integral in (\ref{regint}) the UV and IR  degrees of subtraction --denoted by $\delta$ and $\rho$, respectively, which are defined as follows
\begin{equation*}
\delta=d\,+\,b,\quad \rho=r\,-\,c.
\end{equation*}
$b$ and $c$ are non-negative integers chosen so that
\begin{equation*}
\rho\leq \delta+1.
\end{equation*}
Let us point out that, in our case, since the physical fermion is massless, we always have $d=r$, so that, one can choose $b=0$ and $c=0$ and, thus,
\begin{equation}
\rho=\delta.
\label{rhodelta}
\end{equation}
This is analogous to the situation one encounters when dealing with Yang-Mills theory --see page 154 of  {\cite{Lowenstein:1975ug}. Let us point out that the need to introduce the UV and IR degrees of subtraction is a consequence of the fact --see \cite{Lowenstein:1975ug}-- that our physical fermions are massless. Hence, the divergent UV behaviour of the Feynman integrals is cured by performing appropriate subtractions, governed by $\delta$, at $p_i=0$ --$p_i$ being the external momenta-- and $s=0$. This modifies the right IR behaviour of the Feynman integrals which is restored by suitable IR subtractions, governed by $\rho$, at $p_i=0$ and $s=1$.

Now, let $n$ be an integer. Then, $\tau^n_{x,y}$ applied to a function, $f$, of $x$ and $y$ yields
\begin{equation*}
\begin{array}{l}
{\tau^n_{x,y} f(x,y)=\text{Taylor series of}\, f(x,y)\,\text{at}\, (x,y)=(0,0)\, \text{up to order n}, \quad \text{if}\quad n \geq 0;}\\[4pt]
{\tau^n_{x,y} f(x,y)=0, \quad \text{if}\quad n< 0,}
\end{array}
\end{equation*}
so that the BPHZL subtracted integral, $F^{(sub)}(p,m,M_i,c_i)$, associated to $F[p,m(s-1),c_i,M_i,c_i,\varepsilon]$ in (\ref{regint})  reads
\begin{equation}
F^{(sub)}[p,m,M_i,c_i,\varepsilon] = \idq\;\lim_{s\rightarrow 1}\Big\{(1-\tau^\delta_{p,s-1})(1-\tau^\delta_{p,s}) I[q,p,m(s-1),M_i,c_i,\varepsilon]\Big\}.
\label{subregint}
\end{equation}
Let us indicate that the operator $\tau^\delta_{p,s}$ carries out the UV subtraction mentioned below (\ref{rhodelta}), whereas the operator $\tau^\delta_{p,s-1}$  implements the appropriate IR subtraction --see \cite{Clark:1976ym}.

Now, as shown in \cite{Clark:1976ym},  the limit $M_i\rightarrow\infty$  of  $F^{(sub)}[p,m,M_i,c_i,\varepsilon]$ exists, and, hence, one obtains
the BPHZL renormalized Feynman integral, $F^{(ren)}[p,m,\varepsilon]$, as follows
\begin{equation*}
F^{(ren)}[p,m]=\lim_{\varepsilon\rightarrow 0^{+}}\lim_{M_i\rightarrow \infty}\,F^{(sub)}[p,m,M_i,c_i,\varepsilon],
\end{equation*}
where
\begin{equation}
F^{(ren)}[p,m]=\lim_{\varepsilon\rightarrow 0^{+}}
 \idq\;\lim_{s\rightarrow 1}\Big\{(1-\tau^{\delta-1}_{p,s-1})(1-\tau^\delta_{p,s}) I[q,p,m(s-1),\varepsilon]\Big\}.
\label{Fren}
\end{equation}
$I[q,p,m(s-1),\varepsilon]$ is defined right below (\ref{unregint}).

By applying the renormalization algorithm defined above to each integral over $q$ in section 4, one obtains the one-, two-, three- and four-point contributions to the BPHZL renormalized gravitational effective action. These renormalized  functions will not have parity-odd
terms since the BPHZL subtraction algorithm acts linearly on the integrands and, as we have seen, the cancellation of the parity-odd contributions occur at the integrand level for any values of the regularization parameters and masses. And yet,  the renormalized gravitational effective action so obtained will not be invariant under diffeomorphisms due to the regularization and renormalization processes. However, as is well known (see appendix B), this lack of diffeomorphism invariance can be
removed by adding appropriate UV finite counterterms to the one-, two-, three- and four-point contributions BPHZL renormalized gravitational action. These counterterms will be parity-even since we have shown that the BPHZL renormalized gravitational effective action is parity-even up to
order 4 in the number of $h_{\mu\nu}$'s.

As we have pointed out, the purpose of this paper was not to give close expressions for the renormalized one-, two-, three- and four point contributions to the renormalized
gravitational effective action, but to show that they do not involve any parity-odd contribution. However, we can show that, up to order 4 in the number of  graviton fields, $h_{\mu\nu}$, the diffeomorphism invariant renormalized gravitational effective action constructed as spelled out above is given by half the value of the renormalized effective action for a Dirac spinor non-chirally coupled to gravity in the $MS$-scheme of dimensional regularization, modulo arbitrary diffeomorphism invariant parity-even counterterms. Indeed, the properties of dimensional regularization \cite{Breitenlohner:1975gn, Collins:1984xc} and the fact that $F^{(ren)}[p,m]$, in (\ref{Fren}), is given by an integral which is UV finite by power-counting leads to the conclusion that
\begin{equation*}
F^{(ren)}[p,m]=\lim_{\varepsilon\rightarrow 0^{+}} \lim_{D\rightarrow 4}
 \idqD\;\lim_{s\rightarrow 1}\Big\{(1-\tau^{\delta-1}_{p,s-1})(1-\tau^\delta_{p,s}) I[q,p,m(s-1),\varepsilon]\Big\},
\end{equation*}
where the integral on the right hand side of the latter equation is a dimensionally regularized integral. Next, the linearity property of dimensional regularization implies that
\begin{equation}
\begin{array}{l}
{F^{(ren)}[p,m]=\lim_{\varepsilon\rightarrow 0^{+}} \lim_{D\rightarrow 4}\Big\{
 \idqD\; I[q,p,m(s-1)=0]+}\\[8pt]
 {\idqD\;\lim_{s\rightarrow 1}\Big\{(-\tau^{\delta-1}_{p,s-1}-\tau^\delta_{p,s}+\tau^{\delta-1}_{p,s-1}\tau^\delta_{p,s}) I[q,p,m(s-1)
 ,\varepsilon]\Big\}. }
 \end{array}
 \label{Frendim}
\end{equation}
Now, by construction
\begin{equation*}
{\cal P}[p,m,\varepsilon]=\idqD\;\lim_{s\rightarrow 1}\Big\{(-\tau^{\delta-1}_{p,s-1}-\tau^\delta_{p,s}+\tau^{\delta-1}_{p,s-1}\tau^\delta_{p,s}) I[q,p,m(s-1)
 ,\varepsilon]
\end{equation*}
is a polynomial of degree $\delta$ on the external momenta $p$. The coefficients of this polynomial are Laurent series in  $D-4$, whose singularities are simple poles.
Obviously,
\begin{equation*}
I^{(DR)}[p,\varepsilon]=\idqD\; I[q,p,m(s-1)=0],
\end{equation*}
is the dimensionally regularized integral of the massless theory, which is given in turn by a Laurent series in $D-4$. This series has got a simple pole at $D=4$. As is well known,  the removal of this simple pole yields the corresponding renormalized integral in the $MS$ scheme of dimensional regularization: let us call it $I^{(MS-DR)}[p,\kappa,\varepsilon]$, $\kappa$ being the dimensional regularization scale. Of course the simple pole of ${\cal P}[p,m,\varepsilon]$ at $D=4$ is equal to minus the simple pole of $I^{(DR)}[p,\varepsilon]$, for
$F^{(ren)}[p,m]$ in (\ref{Frendim}) is finite for non-exceptional external momenta. Hence, the difference between $F^{(ren)}[p,m]$ and $I^{(MS-DR)}[p,\kappa,\varepsilon]$ is an UV finite polynomial on the external momenta, i.e., an UV finite local counterterm.

It is plain that the previous analysis can be applied to each parity-even contribution displayed in section 4. Further, the unregularized counterpart of each of these parity-even
contributions is one-half of the corresponding contribution coming  from a Dirac fermion non-chirally coupled to gravity. This is a consequence of the fact that  to obtain such contributions  --in the Pauli-Villars regularization scheme used here-- one moves any projector $P_L=1/2(1-\gamma_5)$ to the far right, and, thus, one ends with a contribution with no $\gamma_5$ and which has the same interaction and propagator structure as the corresponding one  in the theory of a Dirac fermion non-chirally coupled to gravity; but, of course, with an overall $1/2$ coming from the $P_L$.

Finally, putting it all together we conclude that, modulo arbitrary diffeomorphism invariant parity-even counterterms and up to four in the number of $h_{\mu\nu}$'s, the renormalized gravitational effective action of our chiral theory is half the renormalized gravitational effective action of a Dirac spinor non-chirally coupled to gravity.

\section{Using Higher derivative methods.}

The purpose of this section is to show that the results we have presented above can be obtained by using appropriate adaptations of the higher derivative method used in \cite{Nagahama:1985ev} --this method has been taken advantage of to compute the gauge anomaly, and, in \cite{Symanzik:1975rz}, to study certain properties of nonrenormalizable massless $\lambda \phi^4$.

By replacing ${\cal S}_{0}$ in (\ref{action})
with
\begin{equation*}
{\cal S}_{0}^{(HD)}=\,\idx\,\,i\bar{\psi}(x)\gamma^{\mu}\partial_{\mu}\,{\cal K}[\partial^2/\Lambda^2]\psi(x),\quad {\cal K}[\partial^2/\Lambda^2]=\Big[1+\frac{\partial^2}{\Lambda^2}\Big]^n
\end{equation*}
one obtains the following free propagator:
\begin{equation*}
\begin{array}{l}
{\langle \psi_{\alpha}(x)\bar{\psi}_{\beta}(y)\rangle_0^{(HD)}=\idp\; \pslash_{\alpha\beta}\,G^{(HD)}(p)\;e^{-ip(x-y)},}\\[8pt]
{G^{(HD)}(p)=i\,\cfrac{(-\Lambda^{2})^n}{(p^2-\Lambda^2)^n};}
\end{array}
\end{equation*}
which yields a regularized perturbative gravitational effective action if $n\geq 3$.

It is plain that the one- to four-point functions contributions of this effective action can be obtained just by exchanging $G^{(reg)}(p)$ for $G^{(HD)}(p)$ in each Wick contraction in  section 4. Now,  notice that the cancellation mechanism of the  parity-odd contributions displayed in section 4, does not depend on the actual form of $G^{(reg)}(p)$, provided it yields regularized integrals. Hence, it is clear  that all the conclusions we have drawn by using the Pauli-Villars method employed previously remain valid now: no parity-odd contributions and the renormalized gravitational effective action is, up to four graviton fields and modulo arbitrary diffeomorphism invariant parity-even counterterms, equal to half the value of the gravitational effective action of a Dirac field non-chirally coupled to gravity.

\section{Final discussion.}
We have proved that the one-point, two-point, three-point and four-point contributions to the regularized gravitational effective action in (\ref{PVQaction}) carry  no parity-odd contributions. Hence, the BPHZL subtraction of their  UV divergences leads to renormalized parity-even one-point, two-point, three-point and four-point contributions. These renormalized contributions may break diffeomorphism invariance, for the regularization and renormalization processes used  are not invariant under diffeomorphisms.  However, this causes no fundamental problem since invariance under diffeomorphisms is not an anomalous symmetry in four spacetime dimensions and the regularization and renormalization algorithm used here is consistent with the action principle, as discussed in \cite{Clark:1976ym}. Hence, appropriate UV finite counterterms can be introduced to make the new renormalized one-point, two-point, three-point and four point contributions to the effective gravitational action preserve diffeomorphism invariance. These UV finite counterterms are parity-even since the corresponding regularized contributions were parity-even. We finally conclude that by using the formalism put forward here one may define, at least up to order four in the number of $h_{\mu\nu}$'s, a renormalized gravitational effective action which is parity-even and, hence, any breaking of the classical Weyl symmetry which shows  up in the one-point, two-point, three-point and four-point contributions to that effective action is due to parity-even terms, never to parity-odd contributions. This lack of parity-odd contributions in the effective gravitational  action implies that  the Weyl anomaly does not involve parity-odd terms: an outcome that is in complete agreement with the results presented in \cite{Bastianelli:2016nuf, Abdallah:2021eii, Larue:2023tmu, Alvarez:2025mql} and utter disagreement with the result in  \cite{Bonora:2014qla, Bonora:2023soh}.

We also proved in the paper that, up to order four in the graviton field, the renormalized gravitational effective action of our chiral theory can be  defined so that it is given by half the value of the renormalized\footnote{In particular, in the $MS$ scheme of dimensional regularization.} gravitational effective action of a Dirac fermion non-chirally coupled to gravity plus arbitrary diffeomorphism invariant parity-even UV finite counterterms. This in turn implies that the Weyl anomaly of our theory is parity-even and is half the Weyl anomaly which shows up in the theory of a Dirac fermion non-chirally coupled to gravity: this is in agreement with the results
in  \cite{Bastianelli:2016nuf, Abdallah:2021eii, Larue:2023tmu}.

We have also shown that the results presented in the two previous paragraphs can also be obtained by using higher derivative regularization methods.

In view of the computations and results displayed in this paper one may conjecture that there is a renormalization prescription where, at any order in perturbative expansion around Minkowski, the renormalized gravitational effective action of our chiral theory is half the value of the renormalized gravitational effective action of a Dirac fermion non-chirally coupled to the graviton field. This leads to the conclusion that in perturbation theory around Minkowski the partition function, the renormalized partition function of our theory, ${\cal Z}[h_{\mu\nu}]$, is given by
\begin{equation*}
{\cal Z}[h_{\mu\nu}]=\sqrt{\Dirac},
\end{equation*}
where $\Dirac$ denotes the Dirac operator for the perturbative metric $g_{\mu\nu}=\eta_{\mu\nu}+ h_{\mu\nu}$.
Unfortunately, we do not have a proof of the conjecture made, nor whether it is valid for a non flat background and/or at the nonperturbative level. And yet, let us point out that
the results presented in section 4 do not depend on the actual values of $H^{\mu}_{\nu}(x)$ and  $H_{\mu\nu\rho}(x)$ in (\ref{Sintexpan}), so that the contributions to the $n$-point
function of the gravitational effective action coming from them contain no parity-odd terms and, modulo UV finite counterterms, their values are equal to half value of the corresponding contributions computed for a non-chirally coupled Dirac fermion.

We would like to close this section by pointing out that the technique used above to prove the absence of parity-odd contributions we have discussed relies heavily on the properties of the $\gamma_5$ matrix in four dimensions. As has been analyzed in \cite{Breitenlohner:1977hr}, the properties of the  $\gamma_5$ in four dimensions cannot be maintained in the $D$-dimensional space of dimensional regularization. Hence, in the $D$-dimensional space of dimensional regularization, the commutation properties of the $\gamma_5$ with the Dirac matrices must be modified. Up to the best of our knowledge, there is only one mathematically rigorous modification of such commutation properties: it is the one thoroughly formulated in \cite{Breitenlohner:1977hr, Breitenlohner:1976te}. As is well-known this dimensional regularization rigorous framework leads to the occurrence of the so called evanescent operators. It is not clear to us how to apply the algebraic methods employed in this paper to show that these evanescent operators do not give rise to parity-odd contributions to the renormalized three- and four-point terms  in the gravitational effective action discussed in this paper. However, it has been shown in \cite{Abdallah:2021eii} that no parity-odd contributions occur in the corresponding Weyl anomaly, if the rigorous dimensional regularization framework put forward in \cite{Breitenlohner:1977hr, Breitenlohner:1976te} is used. The generalization of the results in \cite{Abdallah:2021eii} to the full renormalized three- and four-point functions is quite a computation and lies outside the scope of this paper. Of course, Physics is independent of the regularization and renormalization schemes.

\section{Acknowledgements.}
We acknowledge stimulating discussions with E. \'Alvarez and L. \'Alvarez-Gaum\'e.  CPM's research work has been financially supported in part by the Spanish Ministry of Science, Innovation and Universities through grant PID2023-149834NB-I00.

\newpage

\appendix

\section{Some detailed computations.}

As instances of the computations which leads to the  results presented in this paper, we shall work out ${\cal T}_1$ and ${\cal T}_2$ in (\ref{triangledefs}) with some detail.

A straightforward application of Wick's theorem leads to
\begin{equation*}
\begin{array}{l}
{{\cal T}_1=-2\int dx_1\,dx_2\,dx_3\,H^{\mu_1}_{1,\nu_1}(x_1)\,H^{\mu_2}_{1,\nu_2}(x_2)H^{\mu_3}_{1,\nu_3}(x_3)\,
(\gamma_{\nu_1}P_L)_{\alpha_1\beta_1}(\gamma_{\nu_2}P_L)_{\alpha_2\beta_2}(\gamma_{\nu_3}P_L)_{\alpha_3\beta_3}
}\\[4pt]
{\phantom{{\cal T}_1=\int dx_1\,dx_2\,}
\langle\partial_{\mu_1}\psi_{\beta_1}(x_1)\bar{\psi}_{\alpha_2}(x_2)\rangle_0\,
\langle\partial_{\mu_2}\psi_{\beta_2}(x_2)\bar{\psi}_{\alpha_3}(x_3)\rangle_0\,\langle\partial_{\mu_3}\psi_{\beta_3}(x_3)\bar{\psi}_{\alpha_1}(x_1
)\rangle_0.}
\end{array}
\end{equation*}
Taking into account the value of the free propagator in (\ref{freeprop}), one gets
\begin{equation*}
\begin{array}{l}
{{\cal T}_1= -2\int\prod\limits_{i=1}^{3}\dpi(2\pi)^4\,\delta(p_1+p_2+p_3)\,H^{\mu_1\nu_1}(p_1)\,H^{\mu_2\nu_2}(p_2)H^{\mu_3\nu_3}(p_3)\Big\{\idq}\\[8pt]
{\Big[
\tr[\gamma_{\nu_1}P_L\gamma_{\sigma_1}\gamma_{\nu_2}P_L\gamma_{\sigma_2}\gamma_{\nu_3}P_L\gamma_{\sigma_3}]\,
q_{\mu_1}q^{\sigma_1}(q+p_2)_{\mu_2}(q+p_2)^{\sigma_2}(q+p_2+p_3)_{\mu_3}(q+p_2+p_3)^{\sigma_3}+}\\[8pt]
{(s-1)\,m\,\tr[\gamma_{\nu_1}P_L\gamma_{\nu_2}P_L\gamma_{\sigma_2}\gamma_{\nu_3}P_L\gamma_{\sigma_3}]\,
q_{\mu_1}(q+p_2)_{\mu_2}(q+p_2)^{\sigma_2}(q+p_2+p_3)_{\mu_3}(q+p_2+p_3)^{\sigma_3}+}\\[8pt]
{(s-1)\,m\,\tr[\gamma_{\nu_1}P_L\gamma_{\sigma_1}\gamma_{\nu_2}P_L\gamma_{\nu_3}P_L\gamma_{\sigma_3}]\,
q_{\mu_1}q^{\sigma_1}(q+p_2)_{\mu_2}(q+p_2+p_3)_{\mu_3}(q+p_2+p_3)^{\sigma_3}+}\\[8pt]
{(s-1)\,m\,\tr[\gamma_{\nu_1}P_L\gamma_{\sigma_1}\gamma_{\nu_2}P_L\gamma_{\sigma_2}\gamma_{\nu_3}P_L]\,
q_{\mu_1}q^{\sigma_1}(q+p_2)_{\mu_2}(q+p_2)^{\sigma_2}(q+p_2+p_3)_{\mu_3}+}\\[8pt]
{(s-1)^2\,m^2\,\tr[\gamma_{\nu_1}P_L\gamma_{\nu_2}P_L\gamma_{\nu_3}P_L\gamma_{\sigma_3}]\,
q_{\mu_1}(q+p_2)_{\mu_2}(q+p_2+p_3)_{\mu_3}(q+p_2+p_3)^{\sigma_3}+}\\[8pt]
{(s-1)^2\,m^2\,\tr[\gamma_{\nu_1}P_L\gamma_{\nu_2}P_L\gamma_{\sigma_2}\gamma_{\nu_3}P_L]\,
q_{\mu_1}(q+p_2)_{\mu_2}(q+p_2)^{\sigma_2}(q+p_2+p_3)_{\mu_3}+}\\[8pt]
{(s-1)^2\,m^2\,\tr[\gamma_{\nu_1}P_L\gamma_{\sigma_1}\gamma_{\nu_2}P_L\gamma_{\nu_3}P_L]\,
q_{\mu_1}(q+p_2)_{\mu_2}(q+p_2+p_3)_{\mu_3}+}\\[8pt]
{(s-1)^3\,m^3\,\tr[\gamma_{\nu_1}P_L\gamma_{\nu_2}P_L\gamma_{\nu_3}P_L]\,
q_{\mu_1}(q+p_2)_{\mu_2}(q+p_2+p_3)_{\mu_3}\Big]}\\[8pt]
{\quad\quad\quad
\,G^{(reg)}(q)\,G^{(reg)}(q+p_2)\,G^{(reg)}(q+p_2+p_3)\Big\}.}
\end{array}
\end{equation*}
Taking into account the last result in (\ref{DiracM}), one concludes that
\begin{equation*}
\begin{array}{l}
{{\cal T}_1= -2\int\prod\limits_{i=1}^{3}\dpi(2\pi)^4\,\delta(p_1\!+\!p_2\!+\!p_3)\,H^{\mu_1\nu_1}(p_1)\,H^{\mu_2\nu_2}(p_2)H^{\mu_3\nu_3}(p_3)\Big\{
\tr[\gamma_{\nu_1}\gamma_{\sigma_1}\gamma_{\nu_2}\gamma_{\sigma_2}\gamma_{\nu_3}\gamma_{\sigma_3}P_R]}\\[8pt]
{\quad\quad\idq\Big[
q_{\mu_1}q^{\sigma_1}(q+p_2)_{\mu_2}(q+p_2)^{\sigma_2}(q+p_2+p_3)_{\mu_3}(q+p_2+p_3)^{\sigma_3}}\\[8pt]
{\phantom{\quad\quad\idq
q_{\mu_1}q^{\sigma_1}(q+p_2)_{\mu_2}(q+p_2)}
 G^{(reg)}(q)\,G^{(reg)}(q+p_2)\,G^{(reg)}(q+p_2+p_3)\Big]\Big\},}
\end{array}
\end{equation*}
where
\begin{equation*}
P_R=\frac{1}{2}\,(1+\gamma_5).
\end{equation*}
Obviously, the previous equation boils down to ${\cal T}_1$ in (\ref{trianglesvalues}).

Next, ${\cal T}_2$ in (\ref{triangledefs}) is given by
\begin{equation*}
\begin{array}{l}
{{\cal T}_2=-2\int dx_1\,dx_2\,dx_3\,H^{\mu_1}_{1,\nu_1}(x_1)\,H^{\mu_2}_{1,\nu_2}(x_2)H^{\mu_3}_{1,\nu_3}(x_3)\,
(\gamma_{\nu_1}P_L)_{\alpha_1\beta_1}(\gamma_{\nu_2}P_L)_{\alpha_2\beta_2}(\gamma_{\nu_3}P_L)_{\alpha_3\beta_3}
}\\[4pt]
{\phantom{{\cal T}_1=\int dx_1\,dx_2\,}
\langle\partial_{\mu_1}\bar{\psi}_{\alpha_1}(x_1)\psi_{\beta_2}(x_2)\rangle_0\,
\langle\partial_{\mu_2}\bar{\psi}_{\alpha_2}(x_2)\psi_{\beta_3}(x_3)\rangle_0\,\langle\partial_{\mu_3}\bar{\psi}_{\alpha_3}(x_3)\psi_{\beta_1}(x_1
)\rangle_0.}
\end{array}
\end{equation*}
By substituting   the regularized  free propagator -.defined in (\ref{freeprop})-- in  the previous equation, one obtains the following result:
\begin{equation*}
\begin{array}{l}
{{\cal T}_2= -2\int\prod\limits_{i=1}^{3}\dpi(2\pi)^4\,\delta(p_1\!+\!p_2\!+\!p_3)\,H^{\mu_1\nu_1}(p_1)\,H^{\mu_2\nu_2}(p_2)H^{\mu_3\nu_3}(p_3)\Big\{
\tr[\gamma_{\sigma_3}\gamma_{\nu_3}\gamma_{\sigma_2}\gamma_{\nu_2}\gamma_{\sigma_1}\gamma_{\nu_1}P_L]}\\[8pt]
{\quad\quad\idq\Big[
q_{\mu_1}q^{\sigma_1}(q+p_2)_{\mu_2}(q+p_2)^{\sigma_2}(q+p_2+p_3)_{\mu_3}(q+p_2+p_3)^{\sigma_3}}\\[8pt]
{\phantom{\quad\quad\idq
q_{\mu_1}q^{\sigma_1}(q+p_2)_{\mu_2}(q+p_2)}
 G^{(reg)}(q)\,G^{(reg)}(q+p_2)\,G^{(reg)}(q+p_2+p_3)\Big]\Big\},}
\end{array}
\end{equation*}
Finally, the  identity
\begin{equation*}
\tr[\gamma_{\sigma_3}\gamma_{\nu_3}\gamma_{\sigma_2}\gamma_{\nu_2}\gamma_{\sigma_1}\gamma_{\nu_1}P_L]=
\tr[\gamma_{\nu_1}\gamma_{\sigma_1}\gamma_{\nu_2}\gamma_{\sigma_2}\gamma_{\nu_3}\gamma_{\sigma_3}P_L]
\end{equation*}
readily leads to ${\cal T}_{2}$ in (\ref{trianglesvalues}).

\section{How to restore diffeomorphism invariance.}

In this appendix we review some well known facts about BRST transformations formally implementing diffeomorphism invariance of the gravitational effective action, the breaking of that invariance through BPHZL renormalization and the restoration of it, in four dimensions,  by introducing UV finite counterterms.

First, let $g_{\mu\nu}=\eta_{\mu\nu}+  h_{\mu\nu}$ be the metric. Then, the BRST transformations induced by infinitesimal diffeomorphisms of the metric read
\begin{equation*}
s\, h_{\mu\nu}=\partial_{\mu}c_{\mu}+\partial_{\nu}c_{\mu}+
\partial_{\mu}c^{\rho}h_{\rho \nu}+\partial_{\nu}c^{\rho}h_{\rho \mu}+c^{\rho}\partial_{\rho}h_{\mu\nu},\quad
s\, c^{\mu}=c^{\rho}\partial_{\rho}c^{\mu}.
\end{equation*}
$c^{\mu}$ denotes the ghost field and $s$ the BRST operator. One can show that $s^2=0$.

The formal effective action ${\cal W}[h_{\mu\nu}]$ in (\ref{Qaction}) in is invariant infinitesimal diffeomorphisms, which is equivalent to
\begin{equation}
s\,{\cal W}[h_{\mu\nu}]=0.
\label{Wardid}
\end{equation}
However, the  BPHZL renormalization process yields a renormalized effective action ${\cal W}^{(BPHZL)}[h_{\mu\nu}]$ which does not satisfy  (\ref{Wardid}),
but
\begin{equation}
s\,{\cal W}^{(BPHZL)}[h_{\mu\nu}]={\cal B}[c^{\mu},h_{\mu\nu}].
\label{BWard}
\end{equation}

Now, since the BPHZL renormalization process satisfies the action principle \cite{Clark:1976ym}, ${\cal B}[c^{\mu},h_{\mu\nu}]$, is a sum of (local) monomials of the fields and their derivatives. By ghost number conservation one concludes that ${\cal B}[c^\mu,h_{\mu\nu}]$  in (\ref{BWard}) is linear in the ghost field and its derivatives.

Next, the fact that there is no pure gravitational anomaly
in four spacetime dimensions \cite{Alvarez-Gaume:1985zzv} leads to the conclusion that
\begin{equation*}
{\cal B}[c^{\mu},h_{\mu\nu}]=s\,{\cal C}[h_{\mu\nu}],
\end{equation*}
where ${\cal C}[h_{\mu\nu}]$ is a sum of monomials of $h_{\mu\nu}$ and its derivatives only: no $c^{\mu}$.

In perturbation theory, ${\cal W}^{(BPHZL)}[h_{\mu\nu}]$ is to be understood as a formal series expansion in the number of fields; hence,
it is necessary to express (\ref{BWard}) in terms of that series expansion. To do so one introduces the linear operators $s_0$ and $s_1$ defined as follows
\begin{equation*}
\begin{array}{l}
{s_0\, h_{\mu\nu}=\partial_{\mu}c_{\mu}+\partial_{\mu}c_{\mu},\quad
s_0\, c^{\mu}=0,}\\[4pt]
{s_1\, h_{\mu\nu}=\partial_{\nu}c^{\rho}h_{\rho \mu}+c^{\rho}\partial_{\rho}h_{\mu\nu},\quad
s_1\, c^{\mu}=c^{\rho}\partial_{\rho}c^{\mu}.}
\end{array}
\end{equation*}
Obviously,
\begin{equation*}
s=s_0+s_1
\end{equation*}

Let us now  express  ${\cal W}^{(BPHZL)}[h_{\mu\nu}]$ and
${\cal C}[h_{\mu\nu}]$ as formal series in the number of
$h_{\mu\nu}$:
\begin{equation*}
{\cal W}^{(BPHZL)}[h_{\mu\nu}]=\sum\limits_{n\geq 1}\,{\cal W}^{(BPHZL)}_n[h_{\mu\nu}],\quad
{\cal C}[h_{\mu\nu}]=\sum\limits_{n\geq 1}\,{\cal C}_n[h_{\mu\nu}],
\end{equation*}
where ${\cal W}^{(BPHZL)}_n$ is the $n$-point contribution to the BPHZL renormalized effective action and
${\cal C}_n$ is a sum of (local) monomials of $h_{\mu\nu}$ and its derivatives of degree $n$. Then, (\ref{BWard}) boils down to the following infinite set of equations
\begin{equation*}
\begin{array}{l}
{s_0\,{\cal W}^{(BPHZL)}_1  =s_0\,{\cal C}_1}\\[4pt]
 {s_0\,{\cal W}^{(BPHZL)}_n+s_1\,{\cal W}^{(BPHZL)}_{n-1}   =s_0\,{\cal C}_n+s_1\,{\cal C}_{n-1},\quad n\geq 2.}
\end{array}
\end{equation*}
Hence, defining a new set of renormalized gravitational $n$-point functions, ${\cal W}^{(dinv)}_n$, as follows
\begin{equation*}
{\cal W}^{(dinv)}_n={\cal W}^{(BPHZL)}_n\,- \, {\cal C}_n,\quad n\geq 1,
\end{equation*}
one concludes that
\begin{equation*}
\begin{array}{l}
{s_0\,{\cal W}^{(dinv)}_1  =0}\\[4pt]
 {s_0\,{\cal W}^{(dinv)}_n+s_1\,{\cal W}^{(dinv)}_{n-1}   =0,\quad n\geq 2.}
\end{array}
\end{equation*}
This means that the new  renormalized gravitational effective action,
\begin{equation*}
{\cal W}^{(dinv)}[h_{\mu\nu}]=\sum\limits_{n\geq 1}\,{\cal W}^{(dinv)}_n[h_{\mu\nu}],
\end{equation*}
is annihilated by the BRST operator $s$, i.e., it is invariant under infinitesimal diffeomorphisms. Recall that in our case ${\cal W}^{(dinv)}_n$, $n\leq 4$, will have no parity-odd terms since we have shown that neither ${\cal W}^{(BPHZL)}_n$ nor ${\cal C}_n$ have them, when $n\leq 4$.

\end{document}